\documentclass[pra,twocolumn,aps,superscriptaddress,showpacs,longbibliography]{revtex4-1}
\usepackage{hyperref}
\hypersetup{
	colorlinks=true,
	citecolor=blue,
	linkcolor=blue,
	urlcolor=blue}
\usepackage{graphicx}
\usepackage[normalem]{ulem}
\usepackage{amsmath}
\usepackage{amsfonts}
\usepackage{amssymb}
\usepackage{epsfig}
\usepackage{subfigure}
\usepackage{mathtools}
\usepackage{braket}
\usepackage[usenames,dvipsnames]{color}
\usepackage{setspace}
\usepackage{bm}
\usepackage{times}
\usepackage{ulem}
\usepackage{dsfont}

\begin{document}
\title[]{Collective excitations in cigar-shaped spin-orbit coupled spin-1 Bose-Einstein condensates}
\author{Rajat}\email{rajat.19phz0009@iitrpr.ac.in}
\affiliation{Department of Physics, Indian Institute of Technology Ropar, Rupnagar-140001, Punjab, India}
\author{Arko Roy}\email{arko@iitmandi.ac.in}
\affiliation{INO-CNR BEC Center and Universit\`a di Trento, via Sommarive 14, I-38123 Trento, Italy}
\affiliation{School of Basic Sciences, Indian Institute of Technology Mandi, Mandi-175075 (H.P.), India}
\author{Sandeep Gautam}\email{sandeep@iitrpr.ac.in}
\affiliation{Department of Physics, Indian Institute of Technology Ropar, Rupnagar-140001, Punjab, India}


\begin{abstract}
We theoretically study the collective excitations of a spin-orbit-coupled spin-1 Bose-Einstein condensate with 
antiferromagnetic spin-exchange interactions in a cigar-shaped trapping potential at zero and finite temperatures
using the Hartree-Fock-Bogoliubov theory with Popov approximation. The collective modes at zero temperature are
corroborated by real-time evolution of the ground state subjected to a perturbation suitable to excite a density
or a spin mode. We have also calculated a few low-lying modes analytically and found a very good agreement with 
the numerical results. We confirm the presence of excitations belonging to two broad categories, namely density,
and spin excitations, based on the calculation of dispersion. The degeneracy between a pair of spin modes is
broken by the spin-orbit coupling. At finite temperature, spin and density excitations show qualitatively
different behavior as a function of temperature.
\end{abstract}
\maketitle
\section{Introduction}
First multi-component Bose-Einstein condensate (BEC) known as spinor condensate was experimentally 
realized with a simultaneous optical trapping of three hyperfine-spin states from $F = 1$ spin manifold in a gas 
of $^{23}$Na atoms \cite{PhysRevLett.80.2027}, and was followed by the demonstration of the condensation with gas of $^{87}$Rb atoms in 
$F = 1$ \cite{PhysRevLett.87.010404} and $F=2$
\cite{PhysRevLett.92.140403,PhysRevA.69.063604,PhysRevLett.92.040402} manifolds. 
The vast literature on the spinor condensates covering both the experimental and theoretical progress has been
reviewed in Refs. \cite{KAWAGUCHI2012253,RevModPhys.85.1191}.
One of the most important advances in the field of cold atom physics in the last decade has been an experimental
demonstration of synthetic spin-orbit (SO) coupling in a pseudospin-half quantum gas~\cite{lin2011} via
light-atom interactions, which has opened up new perspectives in exploring many-body phenomena using
ultracold atoms, such as topological insulators~\cite{RevModPhys.82.3045}, the quantum anomalous Hall 
conductivity ~\cite{PhysRevLett.107.195302} and topological superconductors ~\cite{Zhai_2015}.
In case of spin-1 Bose-Einstein condensates (BECs), SO coupling has also been experimentally realized in a gas 
of $^{87}$Rb~\cite{campbell2016,luo2016} atoms by coupling three hyperfine states with Raman 
lasers, thus paving the way to explore the rich physics of SO-coupled spin-1 BECs. Theoretical predictions and 
experimental observations of SO-coupled spin-1 BECs render various novel ground-state phases including 
plane-wave, stripe or standing-wave, zero-momentum phases, etc.~\cite{PhysRevLett.105.160403,PhysRevLett.117.125301}. 
Distinguishing the different phases close to the phase-transition boundaries through equilibrium density 
profiles is a challenge \cite{PhysRevA.95.033616}. Non-equilibrium transport of spinor BECs entails various
topological excitations such as solitons~\cite{Malomed_2018} and vortices~\cite{Malomed_2018,kasamatsu2005vortices}. To 
characterize the static and dynamical properties of such systems, it is then imperative to study the collective 
excitations manifesting through fluctuations. 

Collective modes, which are the low energy excitations of a quantum gas, can reveal fundamental information
about the ultracold quantum state like the stability of different different ground state phases, fluctuations, 
and superfluidity, etc.~\cite{pethick_smith_2008,pitaevskii2016bose}. To this end, it is experimentally possible 
to excite the low-lying dipole and breathing modes by modulating the harmonic trap and carry out spectroscopic 
studies with utmost precision~\cite{PhysRevLett.77.988}. At zero temperature, the collective excitations of the 
trapless pseudo-spinor and spin-1 BECs with Raman induced SO coupling have been studied 
theoretically~\cite{PhysRevLett.110.235302,PhysRevA.93.033648,PhysRevA.93.023615} and  have been found to exhibit 
roton-maxon structure in zero-momentum and plane-wave phases. Dynamical and energetic instabilities in the 
Raman-induced SO-coupled pseudospinor BEC in a uniform plane-wave phase have also been 
studied~\cite{PhysRevA.87.063610}. Experimental measurement of collective excitations through Bragg 
spectroscopy in Raman induced SO-coupled pseudospinor BECs, revealing the roton-maxon structure, have been 
carried out~\cite{PhysRevA.90.063624,PhysRevLett.114.105301}. Excitation spectrum of a one-dimensional 
quantum droplet for a binary mixture has been examined in Ref.~\cite{PhysRevA.101.051601}. However, on the contrary, theoretical studies of
spinor condensates at finite temperatures are few and demand a thorough investigation.
Experiments usually are performed in harmonic traps and at finite temperatures. It is therefore essential to 
include the effects of trapping potential and fluctuations to theoretically address such systems.
It has been shown that an antiferromagnetic spin-1 Bose gas with fixed norm and magnetization undergoes double 
condensation using Hartree-Fock-Popov \cite{JPSJ.69.3864} and Hartree-Fock theories \cite{PhysRevA.70.043611}. 
Phuc {\it et al.}~\cite{PhysRevA.84.043645} have studied the finite temperature phase diagram of a trapless 
ferromagnetic spin-1 Bose gas in the presence of quadratic Zeeman terms using Hartree-Fock-Bogoliubov theory 
with Popov approximation. Within the framework of Hartree-Fock theory, finite temperature phase diagram of a 
trapless spin-1 BEC with both linear and quadratic Zeeman terms has also been calculated 
\cite{PhysRevA.85.053611}. Experimentally the phase diagram of an antiferromagnetic spin-1 Bose gas has been
studied in Ref.~\cite{PhysRevA.86.061601}. Spin-mixing dynamics of a spin-1 condensate in a highly elongated trap 
has been studied at zero and finite temperature.
\cite{PhysRevA.73.013629,PhysRevLett.99.020404}. The effect of thermal fluctuations on the quantum phase transition
from an antiferromagnetic to phase separated ground state in a spin-1 BEC has been studied
in~\cite{PhysRevA.90.033604}.
The finite temperature phase diagram of a uniform pseudospinor-half BEC with a Raman induced SO coupling shows 
that quantum and thermal fluctuations enlarge the phase space of a plane-wave phase ~\cite{PhysRevA.96.013625}. 
A finite temperature phase transition from a stripe to a plane-wave phase in a Raman induced SO-coupled 
pseudospinor-half $^{87}$Rb Bose gas has been experimentally observed \cite{ji2014experimental}. A perturbation 
approach, valid for small Raman coupling strengths, has been used to study the transition between plane-wave and 
stripe phase for a Raman induced SO-coupled pseudospinor-half BEC at finite temperatures \cite{PhysRevA.90.053608}.
Stability of the plane-wave phase in Rashba SO-coupled pseudospinor condensate with equal intra- and 
inter-species interactions against quantum and thermal fluctuations has been studied ~\cite{PhysRevLett.109.025301}.
Berezinskii-Kosterlitz-Thouless (BKT) superfluid phase transition in an anisotropically SO-coupled 
two-dimensional pseudospinor BEC has been studied using a classical-field Monte Carlo calculations 
~\cite{PhysRevA.95.051601} and stochastic projected Gross-Pitaevskii equation \cite{PhysRevA.95.053629}; the later
study showed the emergence of a true long-range order in the relative phase sector and the quasi-long-range BKT 
order in the total phase sector \cite{PhysRevA.95.053629}.

In this work, we theoretically study the collective excitations of an SO-coupled spin-1 BEC, with 
antiferromagnetic spin-exchange interactions, in a quasi-one-dimensional harmonic trapping potential  
at zero and finite temperature using the Hartree-Fock-Bogoliubov (HFB) theory with Popov approximation. 
To the best of our knowledge, excitation spectra of SO-coupled spin-1 BECs in harmonic trapping potentials
has neither been studied at zero nor at finite temperature. We calculate the collective excitation spectrum 
by numerically solving the generalized GP and Bogoliubov-de-Gennes (BdG) equations self consistently at zero 
and finite temperatures. In addition to this at $T = 0$ K, we also calculate the collective oscillations by 
simulating the real-time propagation of the ground state using $T = 0$ GP equations subjected 
to different kinds of spin and density perturbations.  We also calculate the dispersion 
relation~\cite{PhysRevLett.104.094501} to ascertain the nature of the excitations. To augment our numerical 
results, we use the variational method to calculate the frequencies of a few low-lying modes analytically.

The paper is organized as follows. In Sec. \ref{II}, we describe the HFB theory with Popov approximation for 
an SO coupled spin-1 BEC in a quasi-one-dimensional trapping potential. In Sec. \ref{III-A}, we discuss the 
spectrum of the non-interacting SO-coupled spin-1 BEC. We calculate the collective excitations of SO-coupled 
spin-1 BEC of $^{23}$Na at zero temperature with and without SO coupling by solving the generalized GP and BdG 
equations in a self-consistent manner in Sec. \ref{III-B} and \ref{III-C}. In Sec. \ref{III-D}, we simulate the 
real-time dynamics of suitably perturbed ground state to monitor the dipole and breathing modes
corresponding to density and spin channels, followed by a calculation of a few low-lying modes using variational 
method in Sec. \ref{III-E}. In Sec. \ref{IV}, the excitation spectra at finite temperatures are discussed.   


\section{Model}
\label{II}
We consider a spin-1 SO-coupled spinor BEC in a highly anisotropic harmonic trapping potential
$V(x,y,z) = m \left(\omega_{x}^{2} x^{2}+\omega_{y}^{2} y^{2}+\omega_{z}^{2} 
z^{2}\right)/2$, where $m$ is the atomic mass and $\omega_{y}=\omega_{z}=\omega_{\perp} \gg 
\omega_{x}$. The transverse degrees of freedom are then considered to be frozen, and the system is confined in 
the harmonic oscillator ground state along this direction with $R_x\gg \xi\gg l_{\perp}$, where $R_x$ is the 
half-length of the condensate along $x$ axis, $\xi$ is the density-coherence length, and 
$l_{\perp} = \sqrt{\hbar/m\omega_{\perp}}$. In this case, we can integrate out the $y$ and $z$ coordinates from 
the condensate wave function and describe the system as a quasi-one-dimensional system along the $x$ axis. 
This allows us to consider excitations only along the axial direction $x$. The grand-canonical Hamiltonian in the
second-quantized form for a spin-1 BEC is  $\mathcal{H}=\mathcal{H}_{0}+\mathcal{H}_{\rm int}$, where 
single-particle part of the Hamiltonian $\mathcal{H}_0$ and interaction part of the Hamiltonian 
$\mathcal{H}_{\rm int}$ are \cite{PhysRevLett.81.742,JPSJ.67.1822}
\begin{subequations}\label{ham}
\begin{align}
\mathcal{H}_{0}=\int dx&\hat\psi_{i}^{\dagger}\left[\mathcal{L}_{ij}
                   -\iota \hbar \gamma f_{x}\partial_x\right]\hat\psi_{j},\\
\mathcal{H}_{\rm int}=\int  dx&\left[\frac{c_{0}}{2} \hat\psi_{i}^{\dagger} 
                      \hat\psi_{j}^{\dagger} \hat\psi_{j}    
                      \hat\psi_{i}+\frac{c_{2}}{2} \hat\psi_{i}^{\dagger}(f_{\alpha})_{i j}
                      \hat\psi_{j} \hat\psi_{k}^{\dagger}(f_{\alpha})_{k l} \hat\psi_{l}\right],
\end{align}
\end{subequations}
where $\partial_x = \partial/\partial x$,  $\mathcal{L}_{i j}=\left[(-\hbar^2 /2m)\partial_x^2-\mu+V(x)\right]\delta_{i j}$.
In Eqs.~(\ref{ham}a)-(\ref{ham}b), $i, j, k, l$ which can have values $+1, 0, -1$ 
are the hyperfine spin states of the $F=1$ manifold, repeated indices are summed over, $f_{\alpha}$ with 
$\alpha = x, y, z$ denote the spin-1 matrices in the irreducible representation, 
$\hat\psi_{i}(x,t)\left(\hat\psi_{i}^{\dagger}(x, t)\right)$ is the quantum field for annihilating (creating) 
an atom in state $i$ at position $x$,  $\mu$ is the chemical potential, $\gamma$ is the 
strength of spin-orbit coupling, and $c_0$ and $c_2$ are  spin-independent and spin-dependent
interactions, respectively. The latter expressed in terms of the $s$-wave scattering lengths
$a_0$ and $a_2$ of binary collisions with total spin $F_{\rm total}$ = 0 and 2, respectively, are 
$c_{0} =2\hbar^{2}(a_0 +2 a_2)/(3 m l_{\perp}^{2})$ and $c_{2} =2\hbar^{2}(a_2 - a_0)/(3 m l_{\perp}^{2})$.
Depending on the values of $c_2$, a spin-1 BEC in the absence of SO coupling and Zeeman terms 
admits two phases, namely ferromagnetic for $c_2<0$ and antiferromagnetic for $c_2>0$. 
In the rest of the manuscript, we will work with dimensionless variables (except when stated otherwise) 
defined as $\tilde{x}=x/l_{0}$, $\tilde{E}=E/\hbar \omega_{x}$, $\tilde{t}=\omega_{x}t$, 
$\tilde{\gamma}=\gamma/\sqrt{m\hbar\omega_{x}}$ and
${\tilde c}_{0} =2 (a_0 +2 a_2) l_{0}/(3 l_{\perp}^{2})$,
${\tilde c}_{2} =2 (a_2 - a_0) l_{0}/(3 l_{\perp}^{2})$, 
where $l_0 = \sqrt{\hbar/m\omega_x}$; we further denote the dimensionless variables without tilde
in the rest of the manuscript.

\subsection*{Fluctuations}
To address the effects of quantum and thermal fluctuations in the BECs of dilute atomic gases,
we generalize the HFB theory within the Popov approximation
\cite{PhysRevB.53.9341,PhysRevA.89.013617,PhysRevA.90.023612} and adapt it to an SO-coupled
spin-1 Bose gas. We start with the second-quantized form of the Hamiltonian $\mathcal{H}$ for a
dilute, weakly-interacting Bose gas and derive the generalized GP and the
BdG equations. We separate the Bose field operator $\hat{\psi_i}(x,t)$ 
into a condensate wavefunction $\phi_i(x)$ and a fluctuation operator $\delta{\hat\psi_i}(x,t)$
as $\hat{\psi_i}(x,t)=\phi_i(x)+\delta{\hat\psi_i}(x,t)$.  Employing the  Bogoliubov transformation, 
fluctuation operators $\delta{\hat\psi_i}$ can be expressed as a linear combination of quasi-particle 
creation ($\hat{\alpha}_{\lambda}^{\dagger}$) and annihilation operators ($\hat{\alpha}_{\lambda}$) 
given by
\begin{equation}
\delta{\hat\psi}_{i}(x,t)=\sum_{\lambda}\left[u_{i}^{\lambda}(x) \hat{\alpha}_{\lambda}(x) 
e^{-i \omega_{\lambda} t}-v_{i}^{*\lambda}(x)\hat{\alpha}_{\lambda 
}^{\dagger}(x)e^{i\omega_{\lambda} t}\right],
\end{equation}
where $i \in (+1,0,-1)$ represents the component index, and $\lambda$ represents the eigenvalue
index for the corresponding energy $\omega_{\lambda}$ with $u_{i}^{\lambda}$ and
$v_{i}^{\lambda}$ as the quasiparticle amplitudes of the $i$th component. The quasiparticle 
creation and annihilation operators satisfy the usual Bose commutation relations. 
We consider the Heisenberg equation for the Bose field operator $\hat\psi_i(x, t)$, i.e.,
\begin{equation}
\iota \hbar \frac{\partial}{\partial t} \hat\psi_i(x, t)=[\hat\psi_i(x, t), \mathcal{H}], 
\label{he}
\end{equation}
and then Wick decompose the cubic terms in fluctuation operators as 
$\delta{\hat\psi}^{\dagger}_{i} \delta{\hat\psi}_{j} \delta{\hat\psi}_{k} \simeq
\left\langle\delta{\hat\psi}^{\dagger}_{i} \delta{\hat\psi}_{j}\right\rangle 
\delta{\hat\psi}_{k}+\left\langle\delta{\hat\psi}^{\dagger}_{i} 
\delta{\hat\psi}_{k}\right\rangle \delta{\hat\psi}_{j}+\langle\delta{\hat\psi}_{j}
\delta{\hat\psi}_{k}\rangle \delta{\hat\psi}^{\dagger}_{i}$. We consider the ensemble average
of Eq.~(\ref{he}) and define $n^c_{i}=|\phi_i|^{2},\tilde{n}_{i,j} 
\equiv\langle\delta{\hat\psi}^{\dagger}_{i} \delta{\hat\psi}_{j}\rangle$,
$\tilde{m}_{i,j} \equiv\langle\delta{\hat\psi}_{i} \delta{\hat\psi}_{j}\rangle$ and 
$n_{i} = n^c_{i}+\tilde{n}_{i,i}$ as the local condensate, non-condensate, anomalous and 
total density, respectively, for the $i$th component and $n_{\rm t} = \sum_{i}(n^c_{i}+\tilde{n}_{i,i})$ 
as the total density of the system. To simplify the notation, we denote the thermal density of
the $i$th component $\tilde{n}_{i,i}$ as simply $\tilde{n}_{i}$. The anomalous average terms $\tilde{m}_{i,j}$ are further
neglected to satisfy the Hugenholtz-Pines theorem \cite{PhysRev.116.489}. This forms the
essence of Hartree-Fock-Bogoliubov-Popov approximation. Based on these considerations and
definitions, we arrive at two coupled sets of equations, one for the 
condensate wavefunctions and second for quasiparticle amplitudes.
One set of equations describing the dynamics of the condensate are 
the following coupled generalized GP equations 
\begin{widetext}
\begin{subequations}\label{gpe}
\begin{align}
\iota\frac{\partial \phi_{\pm 1}}{\partial t} &=  [\mathcal{L}_{\pm1,\pm1}+ {c_{0}(n_{\rm 
t}+\tilde{n}_{\pm1,\pm1})}+{c_{2}(n_{\pm1}+n_{0}-n_{\mp1}+\tilde{n}_{\pm1,\pm1})}]\phi_{\pm1}+c_{2} \phi_0^2
\phi_{\mp1}^*\nonumber\\
&+[(c_{0}+c_{2})\tilde{n}_{0,\pm1}+2c_{2}\tilde{n}_{\mp1,0}]\phi_{0}
+(c_{0}-c_{2})\tilde{n}_{\mp1,\pm1}\phi_{\mp1}-i{\frac{\gamma}{\sqrt2}} \partial_x \phi_0,\\
\iota\frac{\partial \phi_0}{\partial t} &= [\mathcal{L}_{0,0}+ c_{0}(n_{\rm 
t}+\tilde{n}_{0,0})+c_{2}(n_{+1}+n_{-1})]\phi_0+2c_{2}\phi_{+1}\phi_0^*\phi_{-1}+[(c_{0}+
c_2)\tilde{n}_{+1,0}+2c_2\tilde{n}_{0,-1}]\phi_{+1}\nonumber\\
&+[(c_{0}+c_2)\tilde{n}_{-1,0}+2c_2\tilde{n}_{0,+1}] 
\phi_{-1}-i \frac{\gamma}{\sqrt{2}}\left(\partial_x \phi_{+1}+ \partial_x \phi_{-1} \right).
\end{align}
\end{subequations}
\end{widetext}


The BdG equations for quasiparticle amplitudes are given as
\begin{equation}
\begin{pmatrix}
M_1 & -M_{2}\\
M^*_2 & -M^*_1
\end{pmatrix} 
\begin{pmatrix}
{\mathbf u}^{\lambda} \\ {\mathbf v}^{\lambda} 
\end{pmatrix} 
=\omega_{\lambda}
\begin{pmatrix}
{\mathbf u}^{\lambda} \\{\mathbf v}^{\lambda} 
\end{pmatrix},
\label{bdg}
\end{equation}
where
\begin{widetext}
\begin{align*}
{M}_{1}&=
\begin{pmatrix}
 h_{+1,+1}+c_{0} n_{+1}+{c_{2} (2 n_{+1}+n_{0}-n_{-1})} & h_{+1,0}+h_{\rm soc} & 
 {(c_{0}-c_{2})(\phi_{-1}^*\phi_{+1}+\tilde{n}_{-1,+1})} \\
h_{+1,0}^*+h_{\rm soc} & h_{0,0}+
c_{0} n_{0}+{c_{2}(n_{+1}+n_{-1})}&h_{0,-1}+h_{\rm soc} \\
 {(c_{0}-c_{2})(\phi_{+1}^*\phi_{-1}+\tilde{n}_{+1,-1})} &  
 h_{0,-1}^*+h_{\rm soc} & 
 h_{-1,-1}+c_{0}n_{-1}+{c_{2}(n_{0}-n_{+1}+2n_{-1})} \nonumber
\end{pmatrix},\\
\end{align*}
\begin{align*}
{M}_{2}=
\begin{pmatrix}
{(c_{0}+c_{2})\phi_{+1}^2} & {(c_{0}+c_{2})\phi_0\phi_{+1}} & 
{(c_{0}-c_{2})\phi_{-1}\phi_{+1}+c_{2}\phi_0^2} \\
{(c_{0}+c_{2})\phi_{+1}\phi_0} &
{c_{0}\phi_{0}^2+2 c_2 \phi_{+1} \phi_{-1}}&
{(c_{0}+c_{2})\phi_{-1}\phi_0} \\
{(c_{0}-c_{2})\phi_{+1}\phi_{-1}+c_{2}\phi_0^2} & {(c_{0}+c_{2})\phi_0\phi_{-1}} & 
(c_{0}+c_{2})\phi_{-1}^{2}
\end{pmatrix},
\end{align*}
\end{widetext}
with
\begin{align}
h_{i, j} =&\left(-\frac{1}{2} \partial_x^2-\mu +V(x)+c_{0}n_{\rm t}\right) 
\delta_{i j},~h_{\rm soc} = {\frac{-\iota\gamma}{\sqrt2}} \partial_x,
\nonumber\\ 
h_{+1,0} &= (c_{0}+c_{2})(\phi_0^*\phi_{+1}+\tilde{n}_{0,+1})+2 c_2 
(\phi_{-1}^*\phi_{0}+\tilde{n}_{-1,0}),\nonumber\\ 
h_{0,-1} &= (c_{0}+c_{2})(\phi_{0}\phi_{-1}^*+\tilde{n}_{-1,0})+2c_2 
(\phi_{+1}\phi_{0}^*+\tilde{n}_{0,+1}),\nonumber\\
{\mathbf u}^{\lambda} &=(u_{+1}^{\lambda}, u_{0}^{\lambda}, u_{-1}^{\lambda})^{\intercal},~ 
{\mathbf v}^{\lambda}=(v_{+1}^{\lambda}, v_{0}^{\lambda}, v_{-1}^{\lambda})^{\intercal}\nonumber,
\end{align}
and ${\intercal}$ denotes the transpose.
The number density of the non-condensate atoms is related to the Bogoliubov quasiparticle
amplitudes through
\begin{eqnarray}
\tilde{n}_{i,j}=& \sum_{\lambda}\left\{\left(u_{i}^{\lambda *} u_{j}^{\lambda} +v_{i}^{\lambda} v_{j}^{\lambda *}\right)
f_{\omega_{\lambda}} \right. \left.+v_{i}^{\lambda} v_{j}^{\lambda *}\right\} \label{td}
\end{eqnarray}
where $ f_{\omega_{\lambda}}=(e^{\omega_{\lambda}/k_{B}T}-1)^{-1}$ is the Bose factor of the $\lambda$th 
quasiparticle state. Furthermore the quasiparticle amplitudes are normalized as
$\int{\sum_{i}(|u_{i}^{\lambda}|^{2}-|v_{i}^{\lambda}|^{2})}dx = 1$, and total number of atoms
is given by $N=\int n_{\rm t} \, dx$. On diagonalizing Eq.~(\ref{bdg}), the energy of the collective
excitations as well as the quasiparticle amplitudes are obtained. These can be used in 
Eq.~(\ref{td}) to obtain the non-condensate densities, which is eventually used in a 
self-consistent computation of Eqs.~(\ref{gpe}a)-(\ref{gpe}b) and (\ref{bdg}) to arrive 
at the condensate and the non-condensate densities. It is to be noted that when $T\rightarrow0$,
the Bose factor $f_{\omega_{\lambda}} \rightarrow 0$, and $\tilde {n}_{i}$ reduces to 
$\sum_{\lambda}|v_{i}^{\lambda}|^{2}$, which accounts for the condensate depletion due to
quantum fluctuations at $T=0$. The stationary ground state solution at $T=0$ is numerically
obtained following Refs.~\cite{KAUR2021107671,banger2021semi} which serves as an initial input for 
computing the non-condensate densities. Using this solution and then discretizing 
Eq.~(\ref{bdg}) by finite difference methods \cite{GAO2020109058}, we cast Eq.~(\ref{bdg})
as a matrix eigenvalue equation and then solve using standard matrix diagonalization algorithms to obtain the 
eigen energies $\omega_{\lambda}$ and quasiparticle amplitudes $u_{i}^{\lambda}$ and
$v_{i}^{\lambda}$. It is to be noted here that we have included the non-condensate density terms, $\tilde{n}_{i,j}$ 
($i\neq j$), in the spin channel in the above equations. These coherence terms between the 
thermal atoms of different components have negligible contribution in scalar BECs but become
important and comparable to the spin-dependent interaction terms in the case of spinor BECs. 
Furthermore, if one does not include these coherence terms, the excitation
spectra for two SO coupling models, namely $H_{\rm SOC}= \gamma p_{x} f_{x}$ and 
$H_{\rm SOC}^{'}= \gamma p_x f_z$, which are related by a rotation about $y$ axis by an angle 
$\pi/2$ turn out to be different at finite as well as at zero temperature in the presence of
quantum fluctuations. The inclusion of $\tilde{n}_{ij}$ with $i\neq j$ renders 
the excitation spectra for the two models equivalent which illustrates the important role
of these terms in accounting for quantum and thermal fluctuations. In fact, even in the absence of SO coupling
if these terms are not included, excitation spectrum of a polar spin-1 BEC is not equivalent to that
of an antiferromagnetic BEC with same interaction strengths in the presence of fluctuations. 
In this context, it is relevant to point out that these terms were not included in several studies on 
spin-1 BECs based on self-consistent solutions of Eqs.~(\ref{gpe}a)-(\ref{gpe}b) and (\ref{bdg}) 
\cite{PhysRevA.70.043611,PhysRevA.104.063308}.

After calculating the collective excitations, one can also compute the associated wavenumbers,
which essentially establishes the dispersion relation. We first compute 
${\tilde u}_{i}^{\lambda}({k})$ and ${\tilde v}_{i}^{\lambda}({k})$, the 
Fourier transforms of the Bogoliubov quasiparticle amplitudes $u_{i}^{\lambda}(x)$ and 
$v_{i}^{\lambda}(x)$, respectively, and then calculate the root-mean-square wave number 
$k_{\rm rms}$ of the $\lambda$th quasiparticle mode  
as \cite{PhysRevLett.104.094501,PhysRevA.89.053601,Pal_2018}
\begin{equation}
k^{\lambda}_{\rm rms}=\sqrt{\frac{\sum_{i} \int dk \, k^{2}
[|{\tilde u}_{i}^{\lambda}(k)|^{2}+|{\tilde v}_{i}^{\lambda}(k)|^{2}]}{\sum_{i} 
\int dk \, [|{\tilde u}_{i}^{\lambda}(k)|^{2}+|{\tilde v}_{i}^{\lambda}(k)|^{2}]}}.
\end{equation}
In Sec.~\ref{III}, we shall demonstrate that density and spin modes have distinct dispersion 
curves. 

\section{Collective excitations at zero temperature}
\label{III}
\subsection{Non-interacting spin-1 BEC}
\label{III-A}
We first analyze the spectrum of the single particle SO-coupled Hamiltonian 
\begin{equation}
H_0 = \mathds{1}\times\left[-\frac{1}{2} \partial_x^2 
      + \frac{x^2}{2} - \mu\right] -\iota \gamma f_x\partial_x,    
\end{equation}
where $\mathds{1}$ is a $3\times3$ identity matrix. As $H_0$ and $U^{\dagger}H_0U$, where $U$ is a unitary
operator, have identical spectra, we consider the $U^{\dagger}H_0U$ with $U$ defined as a rotation operator
which rotates the spin state of a spin-1 particle about $y$ by an angle $\pi/2$ in an anticlockwise direction. 
The unitary operator $U$ is defined as
\begin{equation}
U = 
\begin{pmatrix}
1/2 & -1/\sqrt{2} & 1/2\\
1/\sqrt{2} & 0 & -1/\sqrt{2}\\
1/2 & 1/\sqrt{2} & 1/2
\end{pmatrix}.\label{uni_op}
\end{equation}
The $U^{\dagger}H_0U$ thus obtained in momentum space is
\begin{widetext}
$
\begin{pmatrix}
-\frac{1}{2}\frac{\partial^2}{\partial k^2} + \frac{(k+\gamma)^2}{2} -\frac{\gamma^2}{2}- \mu & 0 & 0\\
0 & -\frac{1}{2}\frac{\partial^2}{\partial k^2} + \frac{k^2}{2} - \mu & 0\\
0 & 0 & -\frac{1}{2}\frac{\partial^2}{\partial k ^2} + \frac{(k-\gamma)^2}{2} -\frac{\gamma^2}{2}- \mu 
\end{pmatrix},
$
\end{widetext}
with eigen functions
\begin{widetext}
\begin{equation}
    \begin{pmatrix}
    \frac{1}{\sqrt{2^n \sqrt{\pi}n!}}\exp\left(\frac{-(k+\gamma)^2}{2}\right)H_n(k)\\
    0\\
    0
    \end{pmatrix},~\begin{pmatrix}
    0\\
    \frac{1}{\sqrt{2^n \sqrt{\pi}n!}}\exp\left(\frac{-k^2}{2}\right)H_n(k)\\
    0
    \end{pmatrix},~\begin{pmatrix}
    0\\
    0\\
    \frac{1}{\sqrt{2^n\sqrt{\pi} n!}}\exp\left(\frac{-(k-\gamma)^2}{2}\right)H_n(k)
    \end{pmatrix}
\end{equation}
\end{widetext}
and eigen spectrum
\begin{equation}
\epsilon_{\pm 1}(n)= \left(n+\frac{1}{2}\right)-\frac{\gamma^2}{2} -\mu,\quad 
\epsilon_{0}(n)= \left(n+\frac{1}{2}\right) -\mu,
\end{equation}
where $n  = 0,1,2,\ldots$, $\mu = (1 - \gamma^2)/2$, and $H_n(k)$ is the Fourier transform of $n$th order
Hermite polynomial. So we get two degenerate
eigen functions and the third is shifted up by $\gamma^2/2$ for each value $n$.
In the absence of a trap, the energy dispersion is
\begin{equation}
\epsilon_{\pm1}(k) = \frac{(k\pm\gamma)^2}{2} -\frac{\gamma^2}{2} - \mu,\quad
\epsilon_{0}(k) = \frac{k^2}{2} - \mu,
\end{equation}
where $\mu$ would be fixed by the number density. The effect of SO coupling is, therefore, to 
open a gap and also shift the minima of $\epsilon_{\pm 1}$ with respect to $\epsilon_{0}$.
The eigen functions of $H_0$ corresponding to eigen energies 
$\epsilon_{\pm 1}(n)$ and $\epsilon_0(n)$ are, respectively,
\begin{align}
    \frac{1}{\sqrt{2^n \sqrt{\pi} n!}}&\exp\left(-\{ x/2 \pm \iota \gamma \}x\right)H_n(x)\begin{pmatrix}
    1/2\\
    \pm 1/\sqrt{2}\\
    1/2
    \end{pmatrix},\\
    \frac{1}{\sqrt{2^n \sqrt{\pi}n!}}&\exp\left(-x^2/2\right)H_n(x)\begin{pmatrix}
    -1/\sqrt{2}\\
    0\\
    1/\sqrt{2}
    \end{pmatrix}.
    \end{align}

\subsection{Interacting spin-1 BEC without SO coupling}
\label{III-B}
\begin{figure}[!hbtp]
\includegraphics[width=\columnwidth]{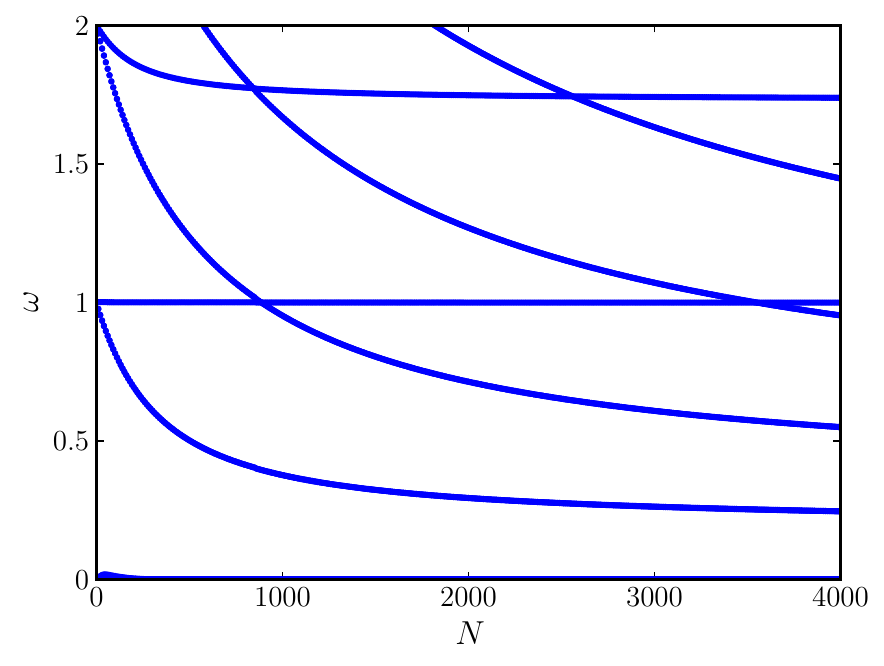}
\caption{Low-lying excitation spectrum for $^{23}$Na spin-1 BEC with ${c_0}=0.0119N$, 
${c_2} = 0.000424N$, and $\gamma = 0$ as a function of number of atoms $N$ at zero temperature. 
Each set of three degenerate modes in the non-interacting limit splits into two branches in the presence of
interactions: the low lying branch consists of two degenerate spin modes and the other corresponds to a single
density mode.}
 \label{eigen-zero-soc}
\end{figure}

We now discuss the role of contact interactions on the ground state of spin-1 spinor condensates at $T=0$ K and 
associated excitation spectra in the absence of spin-orbit coupling and quantum fluctuations. In particular, 
as a representative example, we consider the antiferromagnetic phase of $^{23}$Na~\cite{PhysRevLett.80.2027} 
atoms in the $F=1$ manifold confined in a cigar-shaped trapping potential with the trapping parameters 
$\omega_x=2\pi\times5$ Hz,  $\omega_y=\omega_z=20\omega_x$, and having scattering lengths $a_{0}= 48.91 a_{B}$ 
and $a_{2}=54.54 a_{B}$~\cite{PhysRevA.83.042704}, where $a_{B}$ is the Bohr radius. The interaction strengths, 
in dimensionless units, translate to $c_{0}=0.0119 N$ and $c_{2}=0.000424 N$. With an increase in the number of
atoms, the spatial extent of the ground-state density profiles increases. The antiferromagnetic order constrains
zero population in $m_f=0$ hyperfine state but with an equal population in the $m_f=1$ and $m_f=-1$ states or
all the atoms in $m_f = 0$ state. The 
longitudinal magnetization $M_{z} = \int dx (|\phi_{+1}|^{2}-|\phi_{-1}|^{2})$ is hence equal to zero.
We further investigate the excitation spectrum of the antiferromagnetic phase which is accomplished by
diagonalizing the BdG matrix in (\ref{bdg}). In Fig.~\ref{eigen-zero-soc},
we show the variation in the excitation frequencies with total number of atoms $N$. For a single particle, i.e., when 
the interactions are absent, the excitation spectrum is exactly equivalent to the spectra of three independent harmonic
oscillators. However, in the presence of interactions the equations get coupled. Here the spectrum is characterized 
by the three Goldstone modes with zero excitation frequency. These modes also serve as a self-consistency check 
for the accuracy of our numerical calculations. The presence of three Goldstone modes is attributed to the fact
that for the antiferromagnetic phase, the symmetry group is $U(1)\times S^2$ ~\cite{PhysRevLett.81.742} such that we
can have three broken symmetries. Apart from the one \textit{density} Goldstone mode arising out of the breaking
of $U(1)$ gauge symmetry, the other two \textit{spin} Goldstone modes emanate from the breaking of two symmetry 
generators of the spin rotation~\cite{PhysRevLett.108.251602}. Among the non-zero low-lying modes,
with introduction of the interactions, the degeneracy between the modes with $\omega = 1$ is lifted with a bifurcation into two branches. One branch
corresponds to the density-dipole mode whose energy remains constant with increasing $N$ satisfying the Kohn's 
theorem \cite{PhysRevLett.73.2244}. The other branch consists of two degenerate spin dipole modes. At the 
outset with an increase in $N$, the energy of these spin modes decreases sharply, and then gets saturated for 
higher values of $N$. 
Similarly, three degenerate modes with $\omega = n$ for a non-interacting system where $n = 2, 3, \ldots$
lead to two degenerate spin modes and a density mode having energy higher than corresponding spin modes with
the introduction of interactions as shown in Fig.~\ref{eigen-zero-soc}.
\subsection{Interacting spin-1 BEC with SO coupling}
\label{III-C}
\begin{figure}[!hbtp]
\includegraphics[width=\columnwidth]{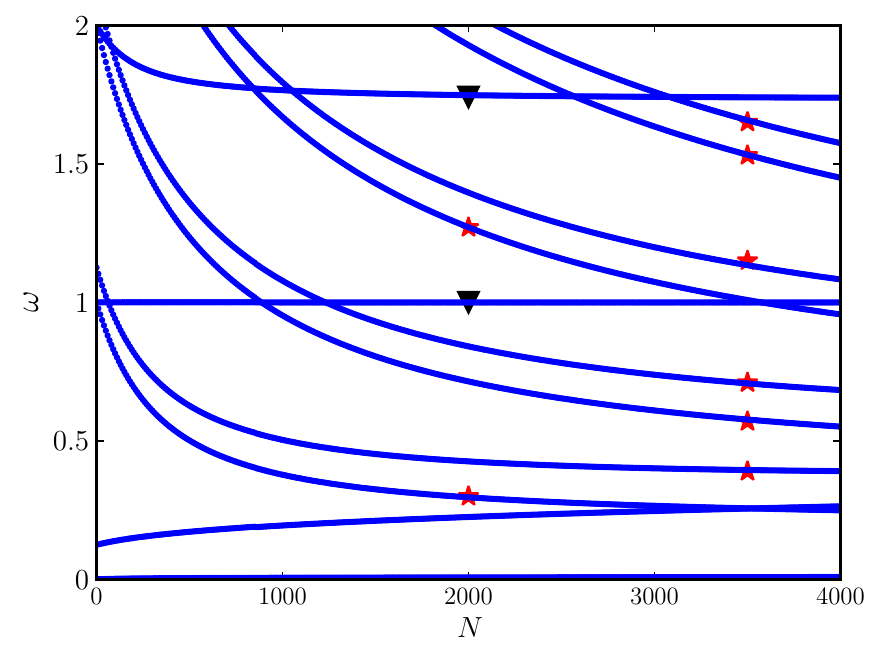}
\caption{Excitation spectrum for $^{23}$Na spin-1 BEC with ${c_0}=0.0119N$, ${c_2} = 0.000424N$ and 
$\gamma = 0.5$ as a function of number of atoms $N$ at zero temperature. The two degenerate modes in the
non-interacting limit, bifurcate with the energy of the spin mode (marked with red stars) becoming lower than the corresponding density
mode (marked with black triangles).}
\label{eigen-0.5-soc}
\end{figure}

\begin{figure}[!hbtp]
\includegraphics[width=\columnwidth]{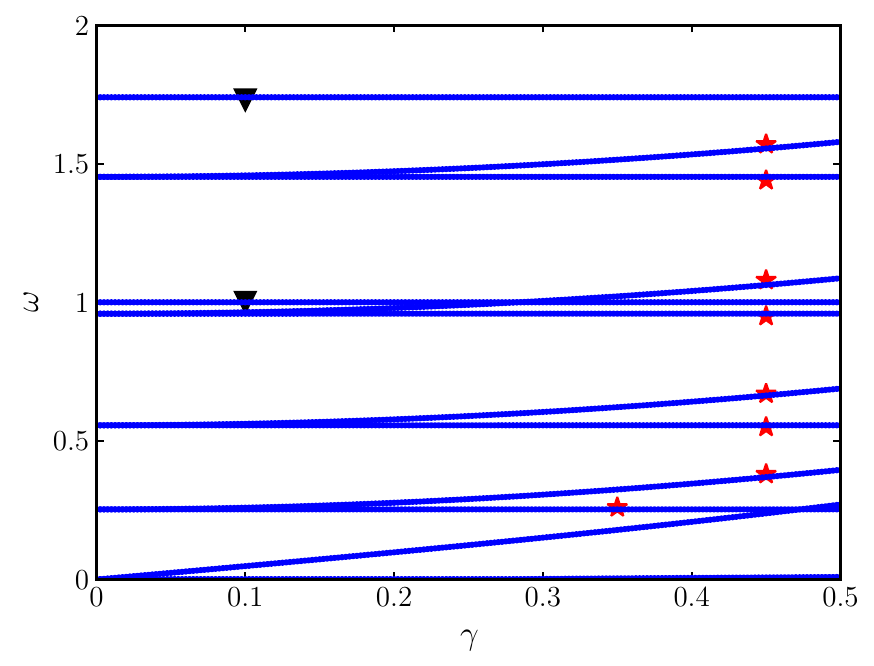}
\caption{Excitation spectrum for $^{23}$Na spin-1 BEC with ${c_0}=0.0119N$, ${c_2} = 0.000424N$ 
and $N$ = 4000 as a function of SO coupling strength $\gamma$ at zero temperature. The modes bifurcating from
$\gamma = 0$ are the spin modes (marked with red stars), whereas the remaining modes not changing with a variation in $\gamma$ are 
density modes (marked with black triangles).}
\label{eigen-with-soc}
\end{figure}

In Fig.~\ref{eigen-0.5-soc}, we show the variation in the excitation frequencies as a function of total number 
of atoms $N$ with fixed SO coupling strength ($\gamma = 0.5$). 
The excitation spectrum is characterized by two zero-energy modes which are identified as two Goldstone modes.
One of these  is a density Goldstone mode arising out of the breaking of $U(1)$ gauge symmetry and the other is 
a spin Goldstone mode originating from the breaking of global $SO(2)$ spin-space rotation 
symmetry~\cite{PhysRevLett.105.160403}. For a small number of atoms, the first non-zero mode is shifted approximately by 
$\gamma^2/2$ as shown in Fig.~\ref{eigen-0.5-soc}. Among the non-zero low lying modes, in the presence of SO 
coupling, is the density-dipole mode, with frequency $\omega = 1$, which remains constant  with increasing $N$
satisfying Kohn's theorem. Another consequence of SO coupling is the lifting of the degeneracy between two spin
modes as can be seen by comparing Figs.~\ref{eigen-zero-soc} and \ref{eigen-0.5-soc}. On increasing the
interaction strengths, the energy of these non-degenerate spin modes decreases, and then gets saturated for
higher values of $N$. The energy separation between these two spin modes remains approximately equal to
$\gamma^2/2$ with a variation in the number of atoms. As expected, for a fixed value of $N$, increasing $\gamma$
increases the energy of one of the spin modes in the pair, whereas the energy of the other spin and density 
modes remain unchanged. This is demonstrated in Fig.~\ref{eigen-with-soc} where excitation spectrum as a 
function of SO-coupling strength is plotted. The modes which are bifurcating from $\gamma=0$ in
Fig.~\ref{eigen-with-soc} are the spin modes, while the remaining modes are the density modes.    
 
\begin{figure}[h]
\includegraphics[width=1.0\columnwidth]{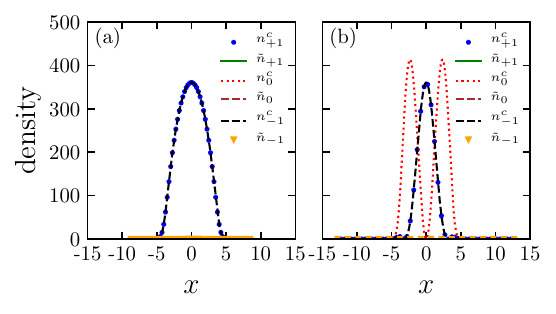}
\caption{(a) Condensate densities, $n^c_{i}(x)$, and thermal densities, $\tilde{n}_i(x)$, for a $^{23}$Na spin-1 
BEC with ${c_0}=0.0119N$, ${c_2}=0.000424N$, $N=4000$, $\gamma = 0.5$ and SO coupling 
$H_{\rm soc}^{'}=\gamma f_z p_x$. The same for $H_{\rm soc} = \gamma f_x p_x$ are shown in (b).}
\label{den_zero_temp}
\end{figure}

To further understand the role of SO coupling on the excitation spectrum, we consider 
$H_{\rm SOC}^{'}=\gamma p_{x} f_{z}$,  where the absence of $m_f = 0$ component as shown in 
Fig.~\ref{den_zero_temp}(a) at $T = 0$ K results in the decoupling of the BdG Eqs.~(\ref{bdg}) corresponding 
to quasiparticle amplitudes $(u_{\pm 1}^\lambda, v_{\pm 1}^\lambda)$ from those for $(u_0^\lambda, v_0^\lambda)$.
On the other hand, for $H_{\rm SOC}=\gamma p_{x} f_{x}$ all $m_f = 0,\pm1$ components are non-zero as shown 
in Fig.~\ref{den_zero_temp}(b)~\cite{PhysRevLett.105.160403}. The solution shown in Fig.~\ref{den_zero_temp}(b) can be obtained by operating
$U$ in  Eq.~(\ref{uni_op}) on the solution shown in Fig.~\ref{den_zero_temp}(a).
The resultant eigen value equation for $(u_{\pm 1}^\lambda, v_{\pm 1}^\lambda)$ with $H_{\rm SOC}^{'}=\gamma p_{x} f_{z}$ is
\begin{widetext}
\begin{equation}
\begin{pmatrix}
A+\iota \gamma \partial_x &-B &C&-D \\
B^{*} &  -A+\iota \gamma \partial_x&D^{*}&C^{*} \\
C^{*}& -D &E-\iota \gamma \partial_x & -F\\
D^{*}&-C&F^{*}&-E-\iota \gamma \partial_x 
\end{pmatrix}
\begin{pmatrix}
u_{+1}^{\lambda} \\v_{+1}^{\lambda} \\u_{-1}^{\lambda} \\v_{-1}^{\lambda}
\end{pmatrix} 
=\omega_{\pm \lambda}
\begin{pmatrix}
u_{+1}^{\lambda} \\v_{+1}^{\lambda} \\u_{-1}^{\lambda} \\v_{-1}^{\lambda}
\end{pmatrix},\label{spin_half_eqs}
\end{equation}
\end{widetext}
where
\begin{align}
A =&\left(-\frac{1}{2}\partial_x^2-\mu +V(x)\right)+2(c_{0}+c_{2})n^c_{+1}\nonumber\\
  &+(c_{0}-{c_{2})n^c_{-1}},\quad B =(c_{0}+c_{2})\phi_{+1}^2,\nonumber\\ 
C =&(c_{0}-c_{2})\phi_{+1}\phi_{-1}^*,\quad D =(c_{0}-c_{2})\phi_{+1}\phi_{-1},\nonumber\\ 
E =&\left(-\frac{1}{2}\partial_x^2-\mu 
    +V(x)\right)+2(c_{0}+c_{2})n^c_{-1}\nonumber\\
   &+(c_{0}-{c_{2})n^c_{+1}},\quad F= (c_{0}+c_{2})\phi_{-1}^2.\nonumber
\end{align}
and that for $(u_0^\lambda,v_0^\lambda)$ is
 \begin{equation}
 \begin{pmatrix}
  R& -S\\
  S^*&
 -R
 \end{pmatrix}
 \begin{pmatrix}
 u_{0}^{\lambda} \\ v_{0}^{\lambda} 
 \end{pmatrix} 
 =\omega_{0\lambda}
 \begin{pmatrix}
 u_{0}^{\lambda} \\v_{0}^{\lambda} 
 \end{pmatrix},\label{bdg_s}
\end{equation}
where $R=\left[-\partial_x^2/2-\mu +V(x)\right]+(c_{0}+c_{2})(n^c_{+1}+n^c_{-1})$ and 
$S =2c_{2}\phi_{+1}\phi_{-1}$. Eq.~(\ref{spin_half_eqs}) is same as the BdG equations of a pseudospinor-1/2
BEC consisting of $m_f = \pm 1$ components with an SO coupling of $\gamma p_x \sigma_z$, where $\sigma_z$ is a 
Pauli spin matrix for spin-1/2 system. Moreover, the eigen modes in (\ref{bdg_s}) are shifted upwards by 
$\gamma^2/2$ compared to the spin mode in  (\ref{spin_half_eqs}). As mentioned earlier, spin-1 BEC with 
$H_{\rm SOC}$ and $H_{\rm SOC}'$ have identical spectra; the quasiparticle amplitudes with the former
can be obtained from a unitary transformation, defined by $U$ in (\ref{uni_op}), of the quasiparticle 
amplitudes for the latter. In Sec. \ref{III-E}, we use the variational method to calculate a few low-lying
modes corresponding to Eq. (\ref{spin_half_eqs}). 

\begin{figure}[!hbtp]
\includegraphics[width=\columnwidth]{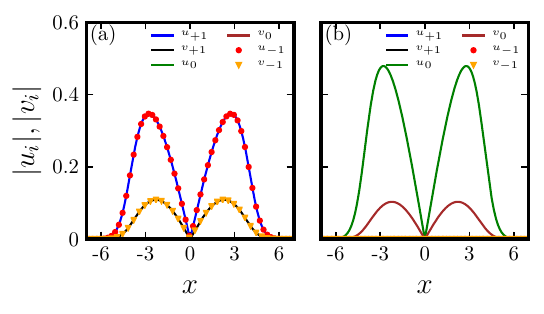}
\caption{(a) and (b) The Bogoliubov amplitudes for the spin-dipole modes with $c_0 = 0.0119N$, $c_2 = 
0.000424N$, $N = 4000$ without and with $H_{\rm SOC}' = \gamma f_z p_x$. Without SO coupling,
(a) and (b) show $|u_i|$ and $|v_i|$s for two degenerate spin-dipole modes. These are also identical to the spin-dipole 
modes in the presence of $H_{\rm SOC}' =0.5f_zp_x$, where (a) and (b) are equivalent to the amplitudes for the spin-dipole modes
with lower and higher frequency, respectively.}
\label{bdg-amplitudes}
\end{figure}

\begin{figure}[!hbtp]
\includegraphics[width=\columnwidth]{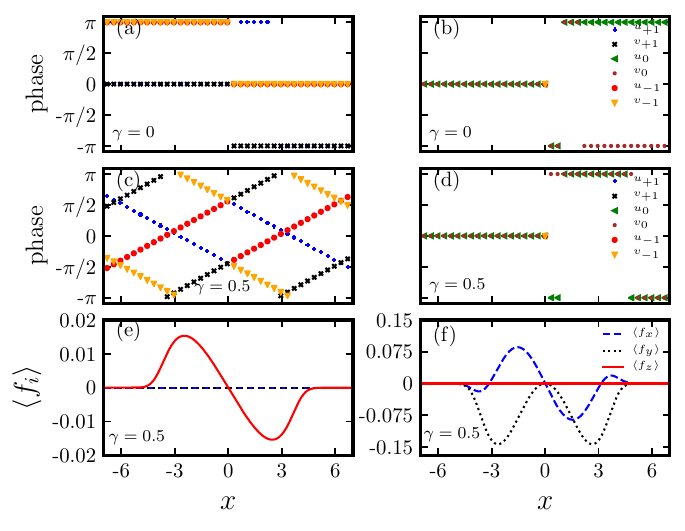} 
\caption{(a), (b) The phase profiles of the spin-dipole modes in the absence of SO coupling corresponding to
the Bogoliubov amplitudes shown in Figs. \ref{bdg-amplitudes}(a), (b). (c) and (d) show the corresponding phase profiles with $H_{\rm SOC}'\ = 0.5f_zp_x$ of 
the lower and higher frequency spin-dipole modes. (e) and (f) are the spin textures corresponding to the lower and higher frequency
spin-dipole modes in the presence of $H_{\rm SOC}'\ = 0.5f_zp_x$. Here $\langle\ldots \rangle$ represents the expectation
with respect to the perturbed order parameter.}
\label{bdg-amp-soc}
\end{figure}

To understand the breakdown in the degeneracy between the pairs of spin modes in the presence of SO
coupling, we analyse their Bogoliubov amplitudes and phases. As an example in Figs.~\ref{bdg-amplitudes}(a) and (b),
the Bogoliubov amplitudes of the two spin-dipole modes for $c_0 = 0.0119N$, $c_2 = 0.000424N$ 
with $N = 4000$ are plotted. The $|u_i|$ and $|v_i|$s for the two modes are identical without and with 
$\gamma f_z p_x$ SO coupling. The excitation frequencies of the modes with $\gamma = 0$ and $\gamma = 0.5$  
are already shown in Fig.~\ref{eigen-zero-soc} and Fig.~\ref{eigen-0.5-soc}, respectively. The phase profiles corresponding 
to the Bogoliubov amplitudes in Figs.~\ref{bdg-amplitudes}(a) and (b) with $\gamma = 0$ are shown in Figs. \ref{bdg-amp-soc}(a) and (b),
respectively, whereas the same with $H_{\rm SOC}' = 0.5 f_z p_x$  are shown in Figs. \ref{bdg-amp-soc}(c) 
and (d), respectively.
With SO coupling the mode with lower excitation frequency acquires SO-coupling 
strength-dependent phase gradient, whereas the phase-profile of the mode with higher excitation frequency remains 
unchanged. Without SO coupling these Bogoliubov modes can be transformed from one to another by a rotation in spin space 
about $y$ axis by an angle $\pi/2$, whereas in the presence of SO coupling the modes are not connected by such a unitary
transformation. The breakdown in degeneracy of the spin-dipole modes is also accompanied by distinct spin-density
vectors (${\bf f} = (\langle f_x \rangle, \langle f_y \rangle, \langle f_z\rangle$) as shown in Figs.~\ref{bdg-amp-soc}(e) and (f) for
lower and higher frequency spin modes, respectively. 

\begin{figure}[!hbtp]
\includegraphics[width=\columnwidth]{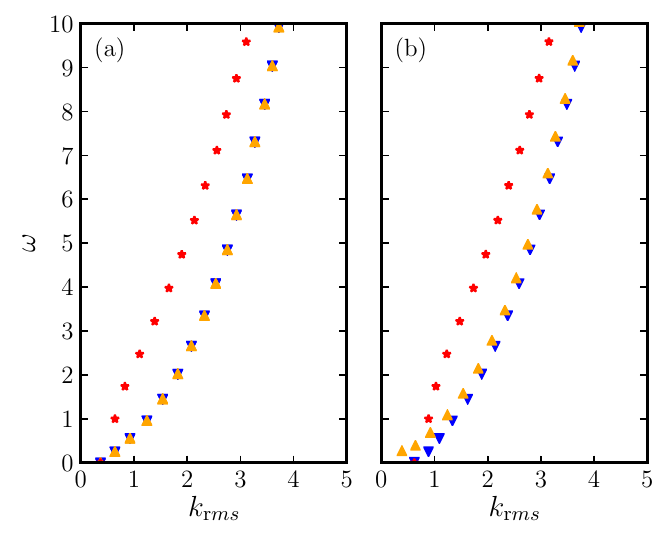}
\caption{(a) Dispersion curves of the $^{23}$Na spin-1 BEC with ${c_0}=0.0119N$, ${c_2} = 0.000424N$, 
$N = 4000$, and $\gamma = 0$. The same for $\gamma = 0.5$ are shown in (b). Red stars represent the dispersion 
for the density modes, whereas upper and lower triangles correspond to dispersion for spin modes. The lifting
of degeneracy between the spin modes in the presence of SO coupling results in three distinct dispersion branches
in (b).}
\label{dispersion-curve}
\end{figure}

We have also numerically computed the dispersion curves for the system as shown in 
Fig.~\ref{dispersion-curve}. For the spin-1 BEC, the spin-independent interaction strength is 
higher than the spin-dependent interaction strength making the energy of density excitations greater 
than the spin excitations for any given $k_{\rm rms}$ as shown in Fig.~\ref{dispersion-curve}(a) in
the absence of SO coupling. In Fig.~\ref{dispersion-curve}(a), the dispersion for the two spin modes 
overlap indicating the degeneracy between the modes. Furthermore, the presence of SO coupling lifts 
the degeneracy between these modes as is shown in Fig.~\ref{dispersion-curve} (b). The dispersion in
the presence and absence of SO coupling are consistent with the excitation spectra in Figs. (\ref{eigen-zero-soc})
and (\ref{eigen-0.5-soc}), respectively.

\subsection {Dynamics}
\label{III-D}
In order to examine the nature of the low-lying collective excitations through physical observables and 
validate our theoretical predictions of mode frequencies obtained from the BdG equations, we perform direct 
numerical dynamical real-time simulations of the system by evolving the ground state with appropriate perturbations. 
The time evolution is done using $T = 0$ K coupled GP equations. 
This type of procedure has already been used experimentally to study low-lying collective excitations by 
modulating the trapping potential for density ~\cite{PhysRevLett.77.988} and spin-dipole 
modes~\cite{PhysRevA.94.063652}. We, however, use perturbation by constructing the fluctuation operator 
corresponding to the dipole and breathing modes in the density and spin channels.  The fluctuation is constructed
with the Bogoliubov quasiparticle amplitudes $u$s and $v$s corresponding to the frequency $\omega$ of the 
relevant mode.  To execute, we add the fluctuation $\delta\psi_i \propto \left(u_i^{\lambda} -v_i^{*\lambda}\right)$ to 
the ground state wave function at time $t=0$ to excite a mode with frequency $\omega_{\lambda}$.
The system is then evolved and a relevant physical observable is monitored over time. We consider $^{23}$Na
spin-1 BEC consisting of $4000$ atoms with $c_0 = 0.0119N$, $c_2 = 0.000434N$ and with a SO coupling $H_{\rm SOC}'= 
\gamma p_x f_z$, where $\gamma = 0.5$ in the remainder of this subsection.

\begin{figure}[!hbtp]
\includegraphics[width=1.0\columnwidth]{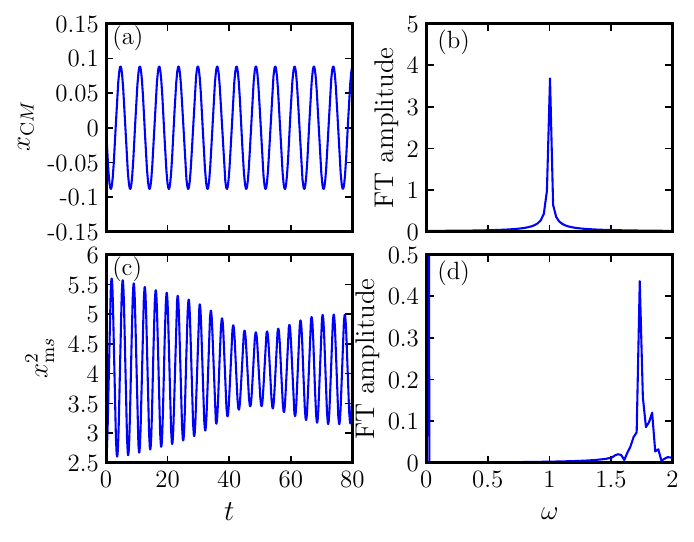}
\caption{(a) shows the center of mass oscillations, i.e, $x_{\rm CM}(t)$ as a function of time and
(b) the Fourier transform of $x_{\rm CM}(t)$ with a dominant peak at $\omega = 1$. Similarly, 
(c) shows the oscillations in the mean square size of the system, $x^2_{\rm ms}(t)$, and
(d) the Fourier transform of $x^2_{\rm ms}(t)$ with a dominant peak at $\omega = 1.73$. The dynamics
corresponds $^{23}$Na spin-1 BEC consisting of $4000$ atoms with $c_0 = 0.0119N$, $c_2=0.000424N$ 
and $\gamma = 0.5$. Both (a) and (c) represent the density excitations.}
\label{db-mode}
\end{figure}

\textit{Density-dipole and breathing modes}: 
To excite the density-dipole mode, we study the center of mass motion 
via $x_{\rm CM}(t) = \langle x \rangle = \sum_{i=-1,0,+1}\int x |\phi_i(x,t)|^2 dx$,
where $\langle\ldots \rangle$ corresponds to an expectation with respect to the time-evolved ground-state. 
In Fig.~\ref{db-mode} (a), we plot
the time dependence of $x_{\rm CM}(t)$. We compute the Fourier transform (FT) of $x_{\rm CM}(t)$ to demonstrate 
that the dominant frequency resonates at $\omega=1$; the FT is shown in Fig.~\ref{db-mode} (b).
Experimentally, the density-dipole mode is excited by a translational shift in the external trapping 
potential~\cite{PhysRevLett.77.988}. Furthermore, to examine the excitation of the breathing mode, 
we consider the corresponding observable $x^2$ and calculate mean square radius 
$x_{\rm ms}^2(t) =  \langle x^2\rangle = \sum_{i=-1,0,+1}\int x^2 |\phi_i(x,t)|^2 dx$ as a function of time. In Fig.~\ref{db-mode} (c) 
and (d), we show the variation in $x^2_{\rm ms}(t)$ and the most dominant peak at $\omega=1.73$ in the
FT of $x^2_{\rm ms}(t)$. 
The density-breathing mode can also be excited by perturbing the trap strength ~\cite{PhysRevLett.77.988}. 
It is worth mentioning here that the frequency of oscillations obtained from the real-time dynamics indeed agrees 
quite well with the corresponding collective excitations obtained from the equilibrium BdG analysis as shown in 
Fig.~\ref{eigen-0.5-soc} for $N=4000$ and $\gamma=0.5$.

\begin{figure}[!hbtp]
\includegraphics[width=1.0\columnwidth]{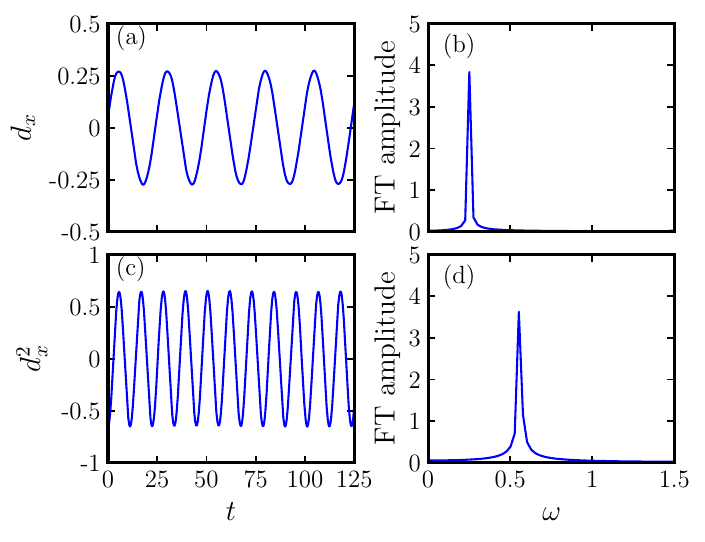}
\caption{Plots showing (a) $d_x(t) = \langle x f_z\rangle$ as a function of time and (b) the Fourier transform of 
$d_x(t)$ with a dominant peak at $\omega=0.25$. (c) and (d) are the same for $d_x^2(t) = \langle x^2 f_z\rangle$
and its Fourier transform with a dominant peak at $\omega=0.55$, respectively. The peaks in (b) and (d) are
the frequencies of spin-dipole and spin-breathing modes, respectively. The dynamics
corresponds $^{23}$Na spin-1 BEC consisting of $4000$ atoms with $c_0 = 0.0119N$, $c_2=0.000424N$ 
and $\gamma = 0.5$. Both (a) and (c) represent the spin excitations.}
\label{sd-sb_mode}
\end{figure}

\begin{figure}[!hbtp]
\includegraphics[width=1.0\columnwidth]{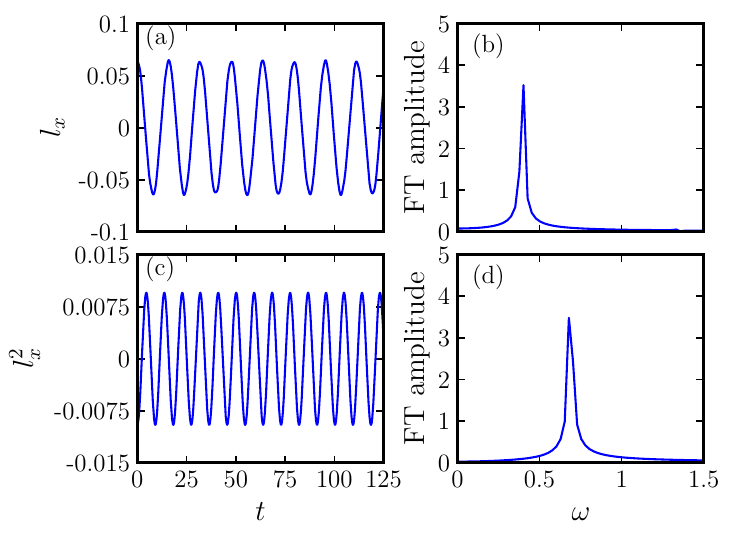}
\caption{Plots showing (a) $l_x(t) = \langle x f_x\rangle$ as a function of time and (b) the Fourier transform 
of $l_x(t)$ with a dominant peak at $\omega=0.395$. (c) and (d) are the same for 
$l_x^2(t) = \langle x^2 f_x\rangle$ and its Fourier transform with a dominant peak at $\omega=0.685$,
respectively. The peaks in (b) and (d) are the frequencies of SSD and SSB modes, respectively. The dynamics
corresponds $^{23}$Na spin-1 BEC consisting of $4000$ atoms with $c_0 = 0.0119N$, $c_2=0.000424N$ and $\gamma = 
0.5$. Both (a) and (c) represent the spin excitations.}
\label{ssd-ssb_mode}
\end{figure}

\textit{Spin-dipole and spin-breathing modes}:
We now turn our attention to excite the spin channel where we first choose the observable 
$x f_{z}$ which corresponds to the spin-dipole mode. We study the dynamics of 
$ d_x(t)= \langle x f_z\rangle = \int \phi_i^*(x,t)(x f_z)_{ij}\phi_{j}(x,t)dx = 
\sum_{i= -1,1} \int x |\phi_{i}(x,t)|^2 dx$. In Fig.~\ref{sd-sb_mode} (a), we plot time dependence of $d_x$ 
and in Fig.~\ref{sd-sb_mode} (b) the frequency dependence of the FT of $d_x$ with a primary peak 
at $\omega=0.25$. Similarly, the spin-breathing mode corresponds to observable $x^2 f_{z}$. 
In Fig.~\ref{sd-sb_mode} (c) and (d), we show the dynamics of $d_x^2 = \langle x^2f_z\rangle=\sum_{i= -1,1} \int x^2 |\phi_{i}(x,t)|^2 dx$,
i.e., the relative difference in the mean-square radii of the $m_f = \pm 1$ components and the associated FT 
with a dominant peak at $\omega=0.55$, respectively. The frequencies of
oscillations thus obtained from the dynamics conform to the BdG analysis of the antiferromagnetic spinor 
condensates at equilibrium as illustrated in Fig.~\ref{eigen-0.5-soc} for $N=4000$ and $\gamma=0.5$.

\textit{Shifted spin-dipole and spin-breathing modes}:
We now discuss the excitation of the spin modes whose energies increase with an increase in SO coupling strength as 
discussed in Sec. \ref{III-C}. As mentioned earlier, these are precisely the modes appearing in Eq.~(\ref{bdg_s}). Here
we first consider the observable $x f_x$ corresponding to the shifted spin-dipole 
mode (SSD) and study the time-dependence of its expectation, i.e., $l_x(t)= \int \phi_i^*(x,t)(x f_x)_{ij}\phi_{j}(x,t)dx$. 
We show $l_x(t)$ as a function of $t$ and the frequency dependence of its FT with a primary peak at
$\omega=0.395$ in Figs.~\ref{ssd-ssb_mode}(a) and (b), respectively. Similarly, the shifted spin-breathing mode (SSB) 
corresponds to the observable $x^2 f_x$. In Fig.~\ref{ssd-ssb_mode} (c) and (d), we illustrate the dynamics of 
$l_x^2(t) = \langle x^2 f_x \rangle=\int \phi_i^*(x,t)(x^2 f_x)_{ij}\phi_{j}(x,t)dx$ and the corresponding FT with a dominant peak 
at $\omega=0.685$, respectively. The frequencies of oscillations thus obtained 
from the dynamics, i.e., 0.395 and 0.685 for SSD and SSB, respectively, agree very well with the BdG analysis of the 
antiferromagnetic spinor condensates at equilibrium as illustrated in Fig.~\ref{eigen-0.5-soc} for $N=4000$ and 
$\gamma=0.5$.  

All of these modes can also be excited for $H_{\rm SOC}=\gamma p_{x} f_{x}$ with exactly the same excitation
frequencies. The relevant observable in this case can 
be obtained by a transformation $U\hat{O}U^\dagger$, where $\hat{O}$ is the observable for $H_{\rm SOC}'=\gamma p_{x} 
f_{z}$. It implies that with $H_{\rm SOC}$, the observable for the density modes will remain the same, i.e., $x$ and 
$x^2$ for density-dipole and density-breathing modes, respectively. On the other hand, the observables for 
spin-dipole and spin-breathing modes are $x f_{x}$ and $x^2 f_{x}$, respectively, while for the shifted spin-dipole and 
spin-breathing modes, operators are $x f_{z}$ and $x^2 f_{z}$, respectively.

\subsection{Variational analysis}
\label{III-E}
For a quasi-one-dimensional spin-1 BEC the spectrum of the low-energy excitations can be studied by using a
time-dependent variational method introduced in Ref. \cite{PhysRevLett.77.5320,PhysRevA.56.1424}. For simplicity, we have considered 
$H_{\rm SOC}^{'}=\gamma p_{x} f_{z}$ which as discussed earlier allows the spectrum analysis as a composition of the spectra of 
two subsystems: one of which corresponds to a pseudospin-1/2 BEC of $m_f = \pm 1$ components with $\gamma p_{x} 
\sigma_{z}$ SO coupling and the second corresponding to the excitation in $m_f = 0$ component of SO-coupled spin-1
BEC. We calculate a 
few low lying modes of the pseudospinor sub-system using the variational method. We consider the Gaussian 
variational {\rm anstaz} 
\begin{align}\label{gauss}
\phi_{\pm1}(x,t)=&A(t) \exp \Bigg[-\frac{\left\{x-x_{\pm1}(t)\right\}^{2}}{2 \sigma(t)^{2}}+\iota \alpha_{\pm1}(t)
\nonumber\\
&\left\{x-x_{\pm1}(t)\right\}+\iota \beta(t) \left\{x-x_{\pm1}(t)\right\}^{2}\Bigg],
\end{align}
where $\sigma$, $x_{\pm1}$, $\alpha_{\pm1}$ and $\beta$ denoting the condensate width, displacement of the $m_f = 
\pm 1$ components from the center of a harmonic trap, phase-gradient and chirp, respectively, are the 
time-dependent variational parameters. The Lagrangian of the subsystem is
\begin{equation}
L=\int \sum_{j=-1,+1} d x \frac{\iota}{2}\left(\phi_{j}^{*} \frac{\partial \phi_{j}}{\partial t}-\phi_{j} \frac{\partial \phi_{j}^{*}}{\partial t}\right)-E
\label{lag}
\end{equation}
where energy $E$ is defined as
\begin{align}
E=& N \int_{-\infty}^{\infty} \Bigg\{\sum_{j=-1,+1}\left(\frac{1}{2}\left|\frac{d \phi_{j}}{d x}\right|^{2}
      +V n^c_{j}\right)\nonumber\\
  &+\frac{c_{0}}{2} (n^c_{+1}+n^c_{-1})^{2} +\frac{c_{2}}{2}\left(n^c_{+1}-n^c_{-1}\right) n^c_{+1}\nonumber\\
  &+\frac{c_{2}}{2}\left(n^c_{-1}-n^c_{+1}\right)  
    n^c_{-1}\nonumber\\
  &+\gamma\left(-\iota \phi_{+1}^{*} \frac{d \phi_{+1}}{d x}+\iota \phi_{-1}^{*} \frac{d \phi_{-1}}{d x}\right)   \Bigg\} d x.\label{En}
\end{align}

We insert Eq.~(\ref{gauss}) in Eq.~(\ref{lag}) and then compute the Euler-Lagrange equations. 
The Euler-Langrange equations for the phase-gradients are $\alpha_{\pm 1}=m \dot{x}_{\pm 1} / \hbar$ and 
for the chirp is $\beta= \dot{\sigma} / 2\sigma$, which are then used in the equation of motion for $x_{\pm 1}$ 
and condensate width $\sigma$ respectively. After linearizing the resultant Euler's equations, 
we get the following equations of motion
\begin{subequations}
\begin{align}
\delta\ddot\sigma(t)+\delta \sigma(t)= -&\sqrt{\frac{2}{\pi}}\frac{c_{0}}{\sigma^{3}}\delta \sigma(t)-\frac{3}{\sigma^{4}}\delta\sigma(t) +  \frac{\sqrt{2}}{\pi} \frac{\left(c_{0}-c_{2}\right)}{\sigma^{5}}\nonumber\\
&\left(x_{+1}-x_{-1}\right)^2 \delta\sigma(t)-\frac{1}{\sqrt{2\pi}} \frac{\left(c_{0}-c_{2}\right)}{\sigma^{4}}\nonumber\\ 
&\left(x_{+1}-x_{-1}\right)
\left(\delta x_{+1}(t)-\delta x_{-1}(t)\right),\\
\delta\ddot{x}_{+1}(t)+\delta x_{+1}(t)&=\frac{1}{2\sqrt{2\pi}}  
                                                 \frac{\left(c_{0}-c_{2}\right)}{\sigma^{3}}
                                                 \left(\delta x_{+1}(t)-\delta x_{-1}(t)\right)\nonumber\\
                                              &-\frac{3}{2\sqrt{2\pi}}\frac{\left(c_{0}-c_{2}\right)}
                                              {\sigma^{4}} \delta \sigma(t)\left( x_{+1}-x_{-1}\right),\\
\delta\ddot{x}_{-1}(t)+\delta x_{-1}(t)&=-\frac{1}{2\sqrt{2\pi}}  
                                                  \frac{\left(c_{0}-c_{2}\right)}{\sigma^{3}}
\left(\delta x_{+1}(t)-\delta x_{-1}(t)\right)\nonumber\\
                                              &+\frac{3}{2\sqrt{2\pi}} 
                                              \frac{\left(c_{0}-c_{2}\right)}{\sigma^{4}} \delta \sigma(t)\left( 
                   x_{+1}-x_{-1}\right).
\end{align}
\label{equ}
\end{subequations}
where $\sigma$ and $x_{\pm 1}$ are to be understood as the equilibrium values. 
For breathing oscillation, we consider
$\delta x_{\pm1}(t) =0$. The equation of motion from (\ref{equ}a) for the condensate width $\sigma$ 
results in
\begin{equation}
 \ddot \delta \sigma(t) + \left(1+ \frac{3}{ \sigma^{4}}+\sqrt{\frac{2}{\pi}}\frac{c_{0}}{\sigma^{3}}\right)\delta\sigma(t) = 0.
\end{equation}
The equilibrium width ($\sigma$) of the condensate satisfies
\begin{equation}
 \sigma^4-\frac{c_{0}\sigma}{\sqrt{2\pi}} = 1,
\end{equation}
and eigen frequency of the oscillations in the width about its equilibrium value is
\begin{equation}
\omega_{b}=\left[1+\frac{3}{ \sigma^{4}}+\sqrt{\frac{2}{\pi}}\frac{c_{0}}{\sigma^{3}}\right]^{1 / 2}
\end{equation}
which is equal to $1.737$ for $c_0 = 0.0119 N$ with $N = 4000$. The variational result matches with 
density-breathing mode of the BdG spectrum shown in Fig.~\ref{eigen-0.5-soc}.
For dipole and spin dipole oscillation, we consider $\delta \sigma(t) = 0$ and add and subtract 
Eqs.~(\ref{equ}b) and (\ref{equ}c) to obtain the following equations of motion
\begin{subequations}
\begin{eqnarray}
\left[\delta\ddot{x}_{+1}(t)+\delta\ddot{x}_{-1}(t)\right]&&+\left[\delta x_{+1}(t)+\delta x_{-1}(t)\right] = 0,\label{dipole}\\
\left[\delta\ddot{x}_{+1}(t)-\delta \ddot{x}_{-1}(t)\right]&&+\left[\delta x_{+1}(t)-\delta x_{-1}(t)\right] = \frac{1}{\sqrt{2\pi}}\nonumber\\ 
&&\frac{\left(c_{0}-c_{2}\right)}{\sigma^{3}}\left[\delta x_{+1}(t)-\delta x_{-1}(t)\right].
\label{sd}
\end{eqnarray}
\end{subequations}
In terms of center of mass coordinate, $\delta x_{1}(t)+\delta x_{-1}(t)$, and relative coordinate, $\delta x_{1}(t)-\delta x_{-1}(t)$, 
Eq.~(\ref{dipole}) corresponds to the center of mass motion oscillating with trap frequency $\omega = 1$ satisfying 
Kohn's theorem and Eq.~(\ref{sd}) corresponds to frequency 
\begin{equation}
\omega_{sd} = \sqrt{\left(1-\sqrt{\frac{1}{2\pi}} \frac{\left(c_{0}-c_{2}\right)}{\sigma^{3}}\right)}
\end {equation}
of spin-dipole mode.
For $c_0 = 0.0119 N$ and $c_2 = 0.000424N$ with $N = 4000$, we get $\omega_{sd}=0.23$ which is very close to 
spin-dipole frequency $0.25$ calculated from BdG analysis shown in Fig.~\ref{eigen-0.5-soc}. Eqs.~(\ref{dipole}), (\ref{sd}) also point out
that in the presence of harmonic trap, center of mass and relative motions are decoupled.
\section{Collective excitations at Finite temperature}
\label{IV}
\begin{figure}
\includegraphics[width=1.0\columnwidth]{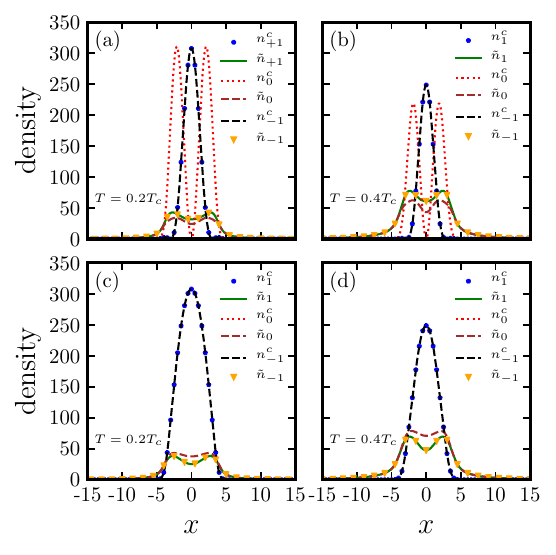}
\caption{Condensate densities, $n^c_{i}(x)$, and thermal densities, $\tilde{n}_{i}(x)$, for $^{23}$Na spin-1 BEC
with ${c_0}=0.00119N$, ${c_2}=0.000424N$, $N=4000$, $\gamma = 0.5$, and $H_{\rm soc}=\gamma f_x p_x$ at (a) $T = 
0.2T_{c}$ and (b) $T = 0.4T_{c}$. The same for for $H^{'}_{\rm soc}=\gamma f_z p_x$ at $T = 0.2T_{c}$ and $T = 
0.4T_{c}$ are shown, respectively, in (c) and (d).}
\label{thermal-den}
\end{figure}

We now analyze the excitation spectra of the SO-coupled $^{23}$Na spin-1 BEC with $c_0 = 0.0119N$, 
$c_2 = 0.000424N$, $\gamma = 0.5$ at finite temperatures and consisting of (a) $N = 2000$ and (b) $N = 4000$. 
The excitation spectrum is calculated by solving Eqs.~(\ref{gpe}a)-(\ref{gpe}b) and ~(\ref{bdg})
self-consistently with the non-condensate density computed from Eq.~(\ref{td}). As an example of static
density profiles, the condensate and the 
non-condensate densities of the system with $N = 4000$ are shown in Fig.~\ref{thermal-den} at $T = 0.2 T_{c}$ 
and $T=0.4 T_{c}$, where $T_{c} = 40.56$ nK is the critical temperature for ideal spin-1 Bose gas in quasi-1D harmonic 
trap~\cite{PhysRevA.54.656,refId0,PhysRevA.65.063610}. 
In Figs.~\ref{thermal-den} (a)-(b), we show the density profiles with $H_{\rm SOC} = \gamma p_x f_x$, 
whereas in (c)-(d) the same with $H_{\rm SOC}'= \gamma p_x f_z$ are shown. 
With increasing temperature, the number of thermal atoms increases along with the spatial extent of the thermal 
cloud. This is accompanied by a corresponding decrease in the condensate density. 
The repulsive interaction between the condensate and non-condensate clouds results in a dip in the 
non-condensate density at the center of the trap and emergence of density peaks towards the edges of the 
trap as shown in Figs.~\ref{thermal-den}(a)-(d).

\begin{figure}
\includegraphics[width=\columnwidth]{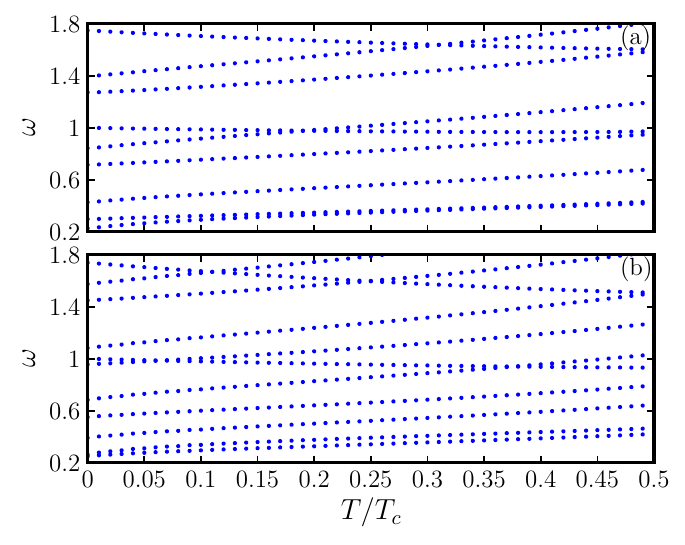}
\caption{(a) The excitation spectrum of $^{23}$Na spin-1 BEC at finite temperature with ${c_0}=0.00119N$, 
${c_2}=0.000424N$, $N=2000$, and $\gamma = 0.5$. The same for $N = 4000$ is shown in (b).}
\label{ft_spectrum}
\end{figure}
In Fig.~\ref{ft_spectrum}(a)-(b), we show the excitation spectra of non-zero modes as a function of temperature
in the presence of SO coupling for $N = 2000$ and $N = 4000$, respectively. With an increase in temperature the
density and the spin modes show qualitatively distinct behavior. We observe that the frequencies of density modes
decrease with an increase in temperature, whereas those of spin modes increase with the temperature. The behavior
of density modes can be understood by the fact that at higher temperatures, the excitations are those of a 
condensate in an effective potential, and that the effective potential is weakened ~\cite{PhysRevA.69.023616} by
the presence of the static thermal cloud, thus lowering the harmonic potential and thereby decreasing the frequency.
On the other hand, the increasing of spin modes' frequencies can be understood by the fact that at finite temperature,
number of atoms in the condensates are decreasing, so by using an equivalent zero temperature condensate (EZC)~\cite{PhysRevA.57.R32}
solution, we get qualitatively same behavior for spin modes as shown in Figs.~\ref{ft_spectrum}. It must be noted
from Fig.~\ref{eigen-0.5-soc} that for $N\ge2000$ the spin modes are much more sensitive to change in number of atoms
in the condensate as compared to the density modes like breathing and dipole modes. The former decrease with an increase in
number of atoms whereas the latter remain almost unchanged. The spin modes therefore increase with an increase of temperature
as the number of atoms in the condensate is decreasing. The density modes on other hand are more sensitive to change in the
effective potential due to the thermal cloud. In order to further ascertain the role of thermal cloud on the
density modes, we extend the variational analysis of Sec. \ref{III-E} to finite temperatures ~\cite{PhysRevA.59.1562} for a
density-breathing mode. We assume the thermal cloud as a static classical gas and write 
~\cite{PhysRevA.59.1562}
\begin{equation}
\tilde{n}_i = r(T) \exp [-V(x) / k_B T],
\end{equation}
where $r(T)$ is the normalization constant. The thermal-cloud density is normalized as $\int \sum_{i} dx  \tilde{n}_i(x)=N_{T}$, where
$N_{T}$ is the number of atoms in thermal cloud. Energy with thermal cloud's contribution included can be written as $E + E_{T}$, where
\begin{eqnarray}
E_{T}&=&[(c_{0}+c_{2})(\tilde{n}_{+1} |\phi_{+1}|^{2}+\tilde{n}_{-1,-1}|\phi_{-1}|^2)\nonumber\\&&+(c_{0}-c_{2})(\tilde{n}_{+1} 
|\phi_{-1}|^{2}+\tilde{n}_{-1,-1}|\phi_{+1}|^2)],\label{E_T}
\end{eqnarray}
and $E$ is defined in Eq.~(\ref{En}). Replacing $E$ by $E+E_T$ in Lagrangian (\ref{lag}) and using the ansatz (\ref{gauss}), the
linearized equation of motion for the condensate width $\sigma$ is
\begin{equation}
 \ddot \delta \sigma(t) + \left(1+ \frac{3}{ \sigma^{4}}+\sqrt{\frac{2}{\pi}}\frac{c_{0}}{\sigma^{3}}-f(T)\right)\delta\sigma(t) = 0,\label{db_ft}
\end{equation}
where $f(T)=\sqrt{\frac{1}{2\pi}}c_{0} N_{T}/(k_{B}T)^{3/2}$ and condensate width
has been considered to be much smaller than the width of the thermal cloud. The frequency of the 
density-breathing mode from Eq.~(\ref{db_ft}) is
\begin{equation}
\omega_{b}=\left[1+\frac{3}{ \sigma^{4}}+\sqrt{\frac{2}{\pi}}\frac{c_{0}}{\sigma^{3}}-f(T)\right]^{1 / 2},
\end{equation}
where $f(T)$ in the square bracket results in the decrease in $\omega_b$ as a function of temperature $T$.
\section{Conclusions}

In this work, we have investigated the collective excitations of a quasi-1D interacting SO-coupled spin-1 BEC 
with antiferromagnetic spin-exchange interactions at zero and finite temperatures by employing the HFB-Popov 
approximation. In this approximation, the static properties of the system are very well described by GP equations
which are coupled to BdG equations. Although as per the Popov approximation, we have neglected the anomalous 
average terms in the generalized GP equations and the BdG equations, we have included and illustrated the 
importance of the coherence terms between the thermal atoms of different components. The low lying modes of the 
interacting spin-1 BEC in the absence of SO coupling are characterized by the three Goldstone modes and doubly
degenerate spin modes. With the introduction of an SO coupling, the degeneracy between the two spin
modes is lifted where one of the modes increases with an increase in SO coupling strength while the other remains 
unchanged. We calculate the dispersion curves showing explicitly that the energy of density excitations are always 
higher than the corresponding spin excitations, and moreover in the presence of SO coupling, dispersion has three distinct
branches consistent with a breakdown in the degeneracy of the spin modes. To substantiate our theoretical 
prediction of the low-lying modes through physical observables useful for experiments, we have performed 
dynamical real-time simulations of the system by evolving the ground state in the presence of perturbations. We 
indeed find the dominant frequencies of oscillation in the center of mass and mean-square radius both in the spin 
and density channels are in excellent agreement with the Bogoliubov calculations. Considering a pseudo-spinor 
subsystem, we have also carried out a time-dependent variational analysis and demonstrated that the analytical 
and numerical results match quite well with each other. The collective excitations and equilibrium density 
profiles at non-zero temperatures are also presented. While the energy of the density modes decreases with an 
increase in temperature, the energy of the spin modes increases. A natural extension of this work would be to 
compute systematically the phase diagram of the interacting SO-coupled spin-1 BEC and to study the dynamics of 
the collective excitations at finite temperatures.

\subsection*{Acknowledgements}
SG acknowledges the support of the Science and Engineering Research Board (SERB), Department of Science and 
Technology, Government of India under the projects ECR/2017/001436 and CRG/2021/002597.

\bibliography{spin_1.bib}{}

\begin{thebibliography}{67}%
\makeatletter
\providecommand \@ifxundefined [1]{%
 \@ifx{#1\undefined}
}%
\providecommand \@ifnum [1]{%
 \ifnum #1\expandafter \@firstoftwo
 \else \expandafter \@secondoftwo
 \fi
}%
\providecommand \@ifx [1]{%
 \ifx #1\expandafter \@firstoftwo
 \else \expandafter \@secondoftwo
 \fi
}%
\providecommand \natexlab [1]{#1}%
\providecommand \enquote  [1]{``#1''}%
\providecommand \bibnamefont  [1]{#1}%
\providecommand \bibfnamefont [1]{#1}%
\providecommand \citenamefont [1]{#1}%
\providecommand \href@noop [0]{\@secondoftwo}%
\providecommand \href [0]{\begingroup \@sanitize@url \@href}%
\providecommand \@href[1]{\@@startlink{#1}\@@href}%
\providecommand \@@href[1]{\endgroup#1\@@endlink}%
\providecommand \@sanitize@url [0]{\catcode `\\12\catcode `\$12\catcode
  `\&12\catcode `\#12\catcode `\^12\catcode `\_12\catcode `\%12\relax}%
\providecommand \@@startlink[1]{}%
\providecommand \@@endlink[0]{}%
\providecommand \url  [0]{\begingroup\@sanitize@url \@url }%
\providecommand \@url [1]{\endgroup\@href {#1}{\urlprefix }}%
\providecommand \urlprefix  [0]{URL }%
\providecommand \Eprint [0]{\href }%
\providecommand \doibase [0]{http://dx.doi.org/}%
\providecommand \selectlanguage [0]{\@gobble}%
\providecommand \bibinfo  [0]{\@secondoftwo}%
\providecommand \bibfield  [0]{\@secondoftwo}%
\providecommand \translation [1]{[#1]}%
\providecommand \BibitemOpen [0]{}%
\providecommand \bibitemStop [0]{}%
\providecommand \bibitemNoStop [0]{.\EOS\space}%
\providecommand \EOS [0]{\spacefactor3000\relax}%
\providecommand \BibitemShut  [1]{\csname bibitem#1\endcsname}%
\let\auto@bib@innerbib\@empty
\bibitem [{\citenamefont {Stamper-Kurn}\ \emph {et~al.}(1998)\citenamefont
  {Stamper-Kurn}, \citenamefont {Andrews}, \citenamefont {Chikkatur},
  \citenamefont {Inouye}, \citenamefont {Miesner}, \citenamefont {Stenger},\
  and\ \citenamefont {Ketterle}}]{PhysRevLett.80.2027}%
  \BibitemOpen
  \bibfield  {author} {\bibinfo {author} {\bibfnamefont {D.~M.}\ \bibnamefont
  {Stamper-Kurn}}, \bibinfo {author} {\bibfnamefont {M.~R.}\ \bibnamefont
  {Andrews}}, \bibinfo {author} {\bibfnamefont {A.~P.}\ \bibnamefont
  {Chikkatur}}, \bibinfo {author} {\bibfnamefont {S.}~\bibnamefont {Inouye}},
  \bibinfo {author} {\bibfnamefont {H.-J.}\ \bibnamefont {Miesner}}, \bibinfo
  {author} {\bibfnamefont {J.}~\bibnamefont {Stenger}}, \ and\ \bibinfo
  {author} {\bibfnamefont {W.}~\bibnamefont {Ketterle}},\ }\href {\doibase
  10.1103/PhysRevLett.80.2027} {\bibfield  {journal} {\bibinfo  {journal}
  {Phys. Rev. Lett.}\ }\textbf {\bibinfo {volume} {80}},\ \bibinfo {pages}
  {2027} (\bibinfo {year} {1998})}\BibitemShut {NoStop}%
\bibitem [{\citenamefont {Barrett}\ \emph {et~al.}(2001)\citenamefont
  {Barrett}, \citenamefont {Sauer},\ and\ \citenamefont
  {Chapman}}]{PhysRevLett.87.010404}%
  \BibitemOpen
  \bibfield  {author} {\bibinfo {author} {\bibfnamefont {M.~D.}\ \bibnamefont
  {Barrett}}, \bibinfo {author} {\bibfnamefont {J.~A.}\ \bibnamefont {Sauer}},
  \ and\ \bibinfo {author} {\bibfnamefont {M.~S.}\ \bibnamefont {Chapman}},\
  }\href {\doibase 10.1103/PhysRevLett.87.010404} {\bibfield  {journal}
  {\bibinfo  {journal} {Phys. Rev. Lett.}\ }\textbf {\bibinfo {volume} {87}},\
  \bibinfo {pages} {010404} (\bibinfo {year} {2001})}\BibitemShut {NoStop}%
\bibitem [{\citenamefont {Chang}\ \emph {et~al.}(2004)\citenamefont {Chang},
  \citenamefont {Hamley}, \citenamefont {Barrett}, \citenamefont {Sauer},
  \citenamefont {Fortier}, \citenamefont {Zhang}, \citenamefont {You},\ and\
  \citenamefont {Chapman}}]{PhysRevLett.92.140403}%
  \BibitemOpen
  \bibfield  {author} {\bibinfo {author} {\bibfnamefont {M.-S.}\ \bibnamefont
  {Chang}}, \bibinfo {author} {\bibfnamefont {C.~D.}\ \bibnamefont {Hamley}},
  \bibinfo {author} {\bibfnamefont {M.~D.}\ \bibnamefont {Barrett}}, \bibinfo
  {author} {\bibfnamefont {J.~A.}\ \bibnamefont {Sauer}}, \bibinfo {author}
  {\bibfnamefont {K.~M.}\ \bibnamefont {Fortier}}, \bibinfo {author}
  {\bibfnamefont {W.}~\bibnamefont {Zhang}}, \bibinfo {author} {\bibfnamefont
  {L.}~\bibnamefont {You}}, \ and\ \bibinfo {author} {\bibfnamefont {M.~S.}\
  \bibnamefont {Chapman}},\ }\href {\doibase 10.1103/PhysRevLett.92.140403}
  {\bibfield  {journal} {\bibinfo  {journal} {Phys. Rev. Lett.}\ }\textbf
  {\bibinfo {volume} {92}},\ \bibinfo {pages} {140403} (\bibinfo {year}
  {2004})}\BibitemShut {NoStop}%
\bibitem [{\citenamefont {Kuwamoto}\ \emph {et~al.}(2004)\citenamefont
  {Kuwamoto}, \citenamefont {Araki}, \citenamefont {Eno},\ and\ \citenamefont
  {Hirano}}]{PhysRevA.69.063604}%
  \BibitemOpen
  \bibfield  {author} {\bibinfo {author} {\bibfnamefont {T.}~\bibnamefont
  {Kuwamoto}}, \bibinfo {author} {\bibfnamefont {K.}~\bibnamefont {Araki}},
  \bibinfo {author} {\bibfnamefont {T.}~\bibnamefont {Eno}}, \ and\ \bibinfo
  {author} {\bibfnamefont {T.}~\bibnamefont {Hirano}},\ }\href {\doibase
  10.1103/PhysRevA.69.063604} {\bibfield  {journal} {\bibinfo  {journal} {Phys.
  Rev. A}\ }\textbf {\bibinfo {volume} {69}},\ \bibinfo {pages} {063604}
  (\bibinfo {year} {2004})}\BibitemShut {NoStop}%
\bibitem [{\citenamefont {Schmaljohann}\ \emph {et~al.}(2004)\citenamefont
  {Schmaljohann}, \citenamefont {Erhard}, \citenamefont {Kronj\"ager},
  \citenamefont {Kottke}, \citenamefont {van Staa}, \citenamefont
  {Cacciapuoti}, \citenamefont {Arlt}, \citenamefont {Bongs},\ and\
  \citenamefont {Sengstock}}]{PhysRevLett.92.040402}%
  \BibitemOpen
  \bibfield  {author} {\bibinfo {author} {\bibfnamefont {H.}~\bibnamefont
  {Schmaljohann}}, \bibinfo {author} {\bibfnamefont {M.}~\bibnamefont
  {Erhard}}, \bibinfo {author} {\bibfnamefont {J.}~\bibnamefont {Kronj\"ager}},
  \bibinfo {author} {\bibfnamefont {M.}~\bibnamefont {Kottke}}, \bibinfo
  {author} {\bibfnamefont {S.}~\bibnamefont {van Staa}}, \bibinfo {author}
  {\bibfnamefont {L.}~\bibnamefont {Cacciapuoti}}, \bibinfo {author}
  {\bibfnamefont {J.~J.}\ \bibnamefont {Arlt}}, \bibinfo {author}
  {\bibfnamefont {K.}~\bibnamefont {Bongs}}, \ and\ \bibinfo {author}
  {\bibfnamefont {K.}~\bibnamefont {Sengstock}},\ }\href {\doibase
  10.1103/PhysRevLett.92.040402} {\bibfield  {journal} {\bibinfo  {journal}
  {Phys. Rev. Lett.}\ }\textbf {\bibinfo {volume} {92}},\ \bibinfo {pages}
  {040402} (\bibinfo {year} {2004})}\BibitemShut {NoStop}%
\bibitem [{\citenamefont {Kawaguchi}\ and\ \citenamefont
  {Ueda}(2012)}]{KAWAGUCHI2012253}%
  \BibitemOpen
  \bibfield  {author} {\bibinfo {author} {\bibfnamefont {Y.}~\bibnamefont
  {Kawaguchi}}\ and\ \bibinfo {author} {\bibfnamefont {M.}~\bibnamefont
  {Ueda}},\ }\href {\doibase https://doi.org/10.1016/j.physrep.2012.07.005}
  {\bibfield  {journal} {\bibinfo  {journal} {Phys. Rep.}\ }\textbf {\bibinfo
  {volume} {520}},\ \bibinfo {pages} {253} (\bibinfo {year} {2012})},\ \bibinfo
  {note} {spinor Bose--Einstein condensates}\BibitemShut {NoStop}%
\bibitem [{\citenamefont {Stamper-Kurn}\ and\ \citenamefont
  {Ueda}(2013)}]{RevModPhys.85.1191}%
  \BibitemOpen
  \bibfield  {author} {\bibinfo {author} {\bibfnamefont {D.~M.}\ \bibnamefont
  {Stamper-Kurn}}\ and\ \bibinfo {author} {\bibfnamefont {M.}~\bibnamefont
  {Ueda}},\ }\href {\doibase 10.1103/RevModPhys.85.1191} {\bibfield  {journal}
  {\bibinfo  {journal} {Rev. Mod. Phys.}\ }\textbf {\bibinfo {volume} {85}},\
  \bibinfo {pages} {1191} (\bibinfo {year} {2013})}\BibitemShut {NoStop}%
\bibitem [{\citenamefont {Lin}\ \emph {et~al.}(2011)\citenamefont {Lin},
  \citenamefont {Jim{\'e}nez-Garc{\'\i}a},\ and\ \citenamefont
  {Spielman}}]{lin2011}%
  \BibitemOpen
  \bibfield  {author} {\bibinfo {author} {\bibfnamefont {Y.-J.}\ \bibnamefont
  {Lin}}, \bibinfo {author} {\bibfnamefont {K.}~\bibnamefont
  {Jim{\'e}nez-Garc{\'\i}a}}, \ and\ \bibinfo {author} {\bibfnamefont {I.~B.}\
  \bibnamefont {Spielman}},\ }\href {\doibase 10.1038/nature09887} {\bibfield
  {journal} {\bibinfo  {journal} {Nature (London)}\ }\textbf {\bibinfo {volume}
  {471}},\ \bibinfo {pages} {83} (\bibinfo {year} {2011})}\BibitemShut
  {NoStop}%
\bibitem [{\citenamefont {Hasan}\ and\ \citenamefont
  {Kane}(2010)}]{RevModPhys.82.3045}%
  \BibitemOpen
  \bibfield  {author} {\bibinfo {author} {\bibfnamefont {M.~Z.}\ \bibnamefont
  {Hasan}}\ and\ \bibinfo {author} {\bibfnamefont {C.~L.}\ \bibnamefont
  {Kane}},\ }\href {\doibase 10.1103/RevModPhys.82.3045} {\bibfield  {journal}
  {\bibinfo  {journal} {Rev. Mod. Phys.}\ }\textbf {\bibinfo {volume} {82}},\
  \bibinfo {pages} {3045} (\bibinfo {year} {2010})}\BibitemShut {NoStop}%
\bibitem [{\citenamefont {van~der Bijl}\ and\ \citenamefont
  {Duine}(2011)}]{PhysRevLett.107.195302}%
  \BibitemOpen
  \bibfield  {author} {\bibinfo {author} {\bibfnamefont {E.}~\bibnamefont
  {van~der Bijl}}\ and\ \bibinfo {author} {\bibfnamefont {R.~A.}\ \bibnamefont
  {Duine}},\ }\href {\doibase 10.1103/PhysRevLett.107.195302} {\bibfield
  {journal} {\bibinfo  {journal} {Phys. Rev. Lett.}\ }\textbf {\bibinfo
  {volume} {107}},\ \bibinfo {pages} {195302} (\bibinfo {year}
  {2011})}\BibitemShut {NoStop}%
\bibitem [{\citenamefont {Zhai}(2015)}]{Zhai_2015}%
  \BibitemOpen
  \bibfield  {author} {\bibinfo {author} {\bibfnamefont {H.}~\bibnamefont
  {Zhai}},\ }\href {\doibase 10.1088/0034-4885/78/2/026001} {\bibfield
  {journal} {\bibinfo  {journal} {Rep. Prog. Phys.}\ }\textbf {\bibinfo
  {volume} {78}},\ \bibinfo {pages} {026001} (\bibinfo {year}
  {2015})}\BibitemShut {NoStop}%
\bibitem [{\citenamefont {Campbell}\ \emph {et~al.}(2016)\citenamefont
  {Campbell}, \citenamefont {Price}, \citenamefont {Putra}, \citenamefont
  {Vald{\'e}s-Curiel}, \citenamefont {Trypogeorgos},\ and\ \citenamefont
  {Spielman}}]{campbell2016}%
  \BibitemOpen
  \bibfield  {author} {\bibinfo {author} {\bibfnamefont {D.}~\bibnamefont
  {Campbell}}, \bibinfo {author} {\bibfnamefont {R.}~\bibnamefont {Price}},
  \bibinfo {author} {\bibfnamefont {A.}~\bibnamefont {Putra}}, \bibinfo
  {author} {\bibfnamefont {A.}~\bibnamefont {Vald{\'e}s-Curiel}}, \bibinfo
  {author} {\bibfnamefont {D.}~\bibnamefont {Trypogeorgos}}, \ and\ \bibinfo
  {author} {\bibfnamefont {I.}~\bibnamefont {Spielman}},\ }\href {\doibase
  10.1038/ncomms10897} {\bibfield  {journal} {\bibinfo  {journal} {Nat.
  Commun.}\ }\textbf {\bibinfo {volume} {7}},\ \bibinfo {pages} {1} (\bibinfo
  {year} {2016})}\BibitemShut {NoStop}%
\bibitem [{\citenamefont {Luo}\ \emph {et~al.}(2016)\citenamefont {Luo},
  \citenamefont {Wu}, \citenamefont {Chen}, \citenamefont {Guan}, \citenamefont
  {Gao}, \citenamefont {Xu}, \citenamefont {You},\ and\ \citenamefont
  {Wang}}]{luo2016}%
  \BibitemOpen
  \bibfield  {author} {\bibinfo {author} {\bibfnamefont {X.}~\bibnamefont
  {Luo}}, \bibinfo {author} {\bibfnamefont {L.}~\bibnamefont {Wu}}, \bibinfo
  {author} {\bibfnamefont {J.}~\bibnamefont {Chen}}, \bibinfo {author}
  {\bibfnamefont {Q.}~\bibnamefont {Guan}}, \bibinfo {author} {\bibfnamefont
  {K.}~\bibnamefont {Gao}}, \bibinfo {author} {\bibfnamefont {Z.-F.}\
  \bibnamefont {Xu}}, \bibinfo {author} {\bibfnamefont {L.}~\bibnamefont
  {You}}, \ and\ \bibinfo {author} {\bibfnamefont {R.}~\bibnamefont {Wang}},\
  }\href {\doibase 10.1038/srep18983} {\bibfield  {journal} {\bibinfo
  {journal} {Sci. Rep.}\ }\textbf {\bibinfo {volume} {6}},\ \bibinfo {pages}
  {1} (\bibinfo {year} {2016})}\BibitemShut {NoStop}%
\bibitem [{\citenamefont {Wang}\ \emph {et~al.}(2010)\citenamefont {Wang},
  \citenamefont {Gao}, \citenamefont {Jian},\ and\ \citenamefont
  {Zhai}}]{PhysRevLett.105.160403}%
  \BibitemOpen
  \bibfield  {author} {\bibinfo {author} {\bibfnamefont {C.}~\bibnamefont
  {Wang}}, \bibinfo {author} {\bibfnamefont {C.}~\bibnamefont {Gao}}, \bibinfo
  {author} {\bibfnamefont {C.-M.}\ \bibnamefont {Jian}}, \ and\ \bibinfo
  {author} {\bibfnamefont {H.}~\bibnamefont {Zhai}},\ }\href {\doibase
  10.1103/PhysRevLett.105.160403} {\bibfield  {journal} {\bibinfo  {journal}
  {Phys. Rev. Lett.}\ }\textbf {\bibinfo {volume} {105}},\ \bibinfo {pages}
  {160403} (\bibinfo {year} {2010})}\BibitemShut {NoStop}%
\bibitem [{\citenamefont {Martone}\ \emph {et~al.}(2016)\citenamefont
  {Martone}, \citenamefont {Pepe}, \citenamefont {Facchi}, \citenamefont
  {Pascazio},\ and\ \citenamefont {Stringari}}]{PhysRevLett.117.125301}%
  \BibitemOpen
  \bibfield  {author} {\bibinfo {author} {\bibfnamefont {G.~I.}\ \bibnamefont
  {Martone}}, \bibinfo {author} {\bibfnamefont {F.~V.}\ \bibnamefont {Pepe}},
  \bibinfo {author} {\bibfnamefont {P.}~\bibnamefont {Facchi}}, \bibinfo
  {author} {\bibfnamefont {S.}~\bibnamefont {Pascazio}}, \ and\ \bibinfo
  {author} {\bibfnamefont {S.}~\bibnamefont {Stringari}},\ }\href {\doibase
  10.1103/PhysRevLett.117.125301} {\bibfield  {journal} {\bibinfo  {journal}
  {Phys. Rev. Lett.}\ }\textbf {\bibinfo {volume} {117}},\ \bibinfo {pages}
  {125301} (\bibinfo {year} {2016})}\BibitemShut {NoStop}%
\bibitem [{\citenamefont {Chen}\ \emph
  {et~al.}(2017{\natexlab{a}})\citenamefont {Chen}, \citenamefont {Pu},
  \citenamefont {Yu},\ and\ \citenamefont {Zhang}}]{PhysRevA.95.033616}%
  \BibitemOpen
  \bibfield  {author} {\bibinfo {author} {\bibfnamefont {L.}~\bibnamefont
  {Chen}}, \bibinfo {author} {\bibfnamefont {H.}~\bibnamefont {Pu}}, \bibinfo
  {author} {\bibfnamefont {Z.-Q.}\ \bibnamefont {Yu}}, \ and\ \bibinfo {author}
  {\bibfnamefont {Y.}~\bibnamefont {Zhang}},\ }\href {\doibase
  10.1103/PhysRevA.95.033616} {\bibfield  {journal} {\bibinfo  {journal} {Phys.
  Rev. A}\ }\textbf {\bibinfo {volume} {95}},\ \bibinfo {pages} {033616}
  (\bibinfo {year} {2017}{\natexlab{a}})}\BibitemShut {NoStop}%
\bibitem [{\citenamefont {Malomed}(2018)}]{Malomed_2018}%
  \BibitemOpen
  \bibfield  {author} {\bibinfo {author} {\bibfnamefont {B.~A.}\ \bibnamefont
  {Malomed}},\ }\href {\doibase 10.1209/0295-5075/122/36001} {\bibfield
  {journal} {\bibinfo  {journal} {Europhys. Lett.}\ }\textbf {\bibinfo {volume}
  {122}},\ \bibinfo {pages} {36001} (\bibinfo {year} {2018})}\BibitemShut
  {NoStop}%
\bibitem [{\citenamefont {Kasamatsu}\ \emph {et~al.}(2005)\citenamefont
  {Kasamatsu}, \citenamefont {Tsubota},\ and\ \citenamefont
  {Ueda}}]{kasamatsu2005vortices}%
  \BibitemOpen
  \bibfield  {author} {\bibinfo {author} {\bibfnamefont {K.}~\bibnamefont
  {Kasamatsu}}, \bibinfo {author} {\bibfnamefont {M.}~\bibnamefont {Tsubota}},
  \ and\ \bibinfo {author} {\bibfnamefont {M.}~\bibnamefont {Ueda}},\ }\href
  {\doibase 10.1142/S0217979205029602} {\bibfield  {journal} {\bibinfo
  {journal} {Int. J. Mod. Phys. B}\ }\textbf {\bibinfo {volume} {19}},\
  \bibinfo {pages} {1835} (\bibinfo {year} {2005})}\BibitemShut {NoStop}%
\bibitem [{\citenamefont {Pethick}\ and\ \citenamefont
  {Smith}(2008)}]{pethick_smith_2008}%
  \BibitemOpen
  \bibfield  {author} {\bibinfo {author} {\bibfnamefont {C.~J.}\ \bibnamefont
  {Pethick}}\ and\ \bibinfo {author} {\bibfnamefont {H.}~\bibnamefont
  {Smith}},\ }\href {\doibase https://doi.org/10.1017/CBO9780511802850} {\emph
  {\bibinfo {title} {Bose–Einstein Condensation in Dilute Gases}}}\ (\bibinfo
   {publisher} {Cambridge University Press},\ \bibinfo {year}
  {2008})\BibitemShut {NoStop}%
\bibitem [{\citenamefont {Pitaevskii}\ and\ \citenamefont
  {Stringari}(2016)}]{pitaevskii2016bose}%
  \BibitemOpen
  \bibfield  {author} {\bibinfo {author} {\bibfnamefont {L.}~\bibnamefont
  {Pitaevskii}}\ and\ \bibinfo {author} {\bibfnamefont {S.}~\bibnamefont
  {Stringari}},\ }\href {\doibase 10.1093/acprof:oso/9780198758884.001.0001}
  {\emph {\bibinfo {title} {Bose-Einstein condensation and superfluidity}}},\
  Vol.\ \bibinfo {volume} {164}\ (\bibinfo  {publisher} {Oxford University
  Press},\ \bibinfo {year} {2016})\BibitemShut {NoStop}%
\bibitem [{\citenamefont {Mewes}\ \emph {et~al.}(1996)\citenamefont {Mewes},
  \citenamefont {Andrews}, \citenamefont {van Druten}, \citenamefont {Kurn},
  \citenamefont {Durfee}, \citenamefont {Townsend},\ and\ \citenamefont
  {Ketterle}}]{PhysRevLett.77.988}%
  \BibitemOpen
  \bibfield  {author} {\bibinfo {author} {\bibfnamefont {M.-O.}\ \bibnamefont
  {Mewes}}, \bibinfo {author} {\bibfnamefont {M.~R.}\ \bibnamefont {Andrews}},
  \bibinfo {author} {\bibfnamefont {N.~J.}\ \bibnamefont {van Druten}},
  \bibinfo {author} {\bibfnamefont {D.~M.}\ \bibnamefont {Kurn}}, \bibinfo
  {author} {\bibfnamefont {D.~S.}\ \bibnamefont {Durfee}}, \bibinfo {author}
  {\bibfnamefont {C.~G.}\ \bibnamefont {Townsend}}, \ and\ \bibinfo {author}
  {\bibfnamefont {W.}~\bibnamefont {Ketterle}},\ }\href {\doibase
  10.1103/PhysRevLett.77.988} {\bibfield  {journal} {\bibinfo  {journal} {Phys.
  Rev. Lett.}\ }\textbf {\bibinfo {volume} {77}},\ \bibinfo {pages} {988}
  (\bibinfo {year} {1996})}\BibitemShut {NoStop}%
\bibitem [{\citenamefont {Li}\ \emph {et~al.}(2013)\citenamefont {Li},
  \citenamefont {Martone}, \citenamefont {Pitaevskii},\ and\ \citenamefont
  {Stringari}}]{PhysRevLett.110.235302}%
  \BibitemOpen
  \bibfield  {author} {\bibinfo {author} {\bibfnamefont {Y.}~\bibnamefont
  {Li}}, \bibinfo {author} {\bibfnamefont {G.~I.}\ \bibnamefont {Martone}},
  \bibinfo {author} {\bibfnamefont {L.~P.}\ \bibnamefont {Pitaevskii}}, \ and\
  \bibinfo {author} {\bibfnamefont {S.}~\bibnamefont {Stringari}},\ }\href
  {\doibase 10.1103/PhysRevLett.110.235302} {\bibfield  {journal} {\bibinfo
  {journal} {Phys. Rev. Lett.}\ }\textbf {\bibinfo {volume} {110}},\ \bibinfo
  {pages} {235302} (\bibinfo {year} {2013})}\BibitemShut {NoStop}%
\bibitem [{\citenamefont {Yu}(2016)}]{PhysRevA.93.033648}%
  \BibitemOpen
  \bibfield  {author} {\bibinfo {author} {\bibfnamefont {Z.-Q.}\ \bibnamefont
  {Yu}},\ }\href {\doibase 10.1103/PhysRevA.93.033648} {\bibfield  {journal}
  {\bibinfo  {journal} {Phys. Rev. A}\ }\textbf {\bibinfo {volume} {93}},\
  \bibinfo {pages} {033648} (\bibinfo {year} {2016})}\BibitemShut {NoStop}%
\bibitem [{\citenamefont {Sun}\ \emph {et~al.}(2016)\citenamefont {Sun},
  \citenamefont {Qu}, \citenamefont {Xu}, \citenamefont {Zhang},\ and\
  \citenamefont {Zhang}}]{PhysRevA.93.023615}%
  \BibitemOpen
  \bibfield  {author} {\bibinfo {author} {\bibfnamefont {K.}~\bibnamefont
  {Sun}}, \bibinfo {author} {\bibfnamefont {C.}~\bibnamefont {Qu}}, \bibinfo
  {author} {\bibfnamefont {Y.}~\bibnamefont {Xu}}, \bibinfo {author}
  {\bibfnamefont {Y.}~\bibnamefont {Zhang}}, \ and\ \bibinfo {author}
  {\bibfnamefont {C.}~\bibnamefont {Zhang}},\ }\href {\doibase
  10.1103/PhysRevA.93.023615} {\bibfield  {journal} {\bibinfo  {journal} {Phys.
  Rev. A}\ }\textbf {\bibinfo {volume} {93}},\ \bibinfo {pages} {023615}
  (\bibinfo {year} {2016})}\BibitemShut {NoStop}%
\bibitem [{\citenamefont {Ozawa}\ \emph {et~al.}(2013)\citenamefont {Ozawa},
  \citenamefont {Pitaevskii},\ and\ \citenamefont
  {Stringari}}]{PhysRevA.87.063610}%
  \BibitemOpen
  \bibfield  {author} {\bibinfo {author} {\bibfnamefont {T.}~\bibnamefont
  {Ozawa}}, \bibinfo {author} {\bibfnamefont {L.~P.}\ \bibnamefont
  {Pitaevskii}}, \ and\ \bibinfo {author} {\bibfnamefont {S.}~\bibnamefont
  {Stringari}},\ }\href {\doibase 10.1103/PhysRevA.87.063610} {\bibfield
  {journal} {\bibinfo  {journal} {Phys. Rev. A}\ }\textbf {\bibinfo {volume}
  {87}},\ \bibinfo {pages} {063610} (\bibinfo {year} {2013})}\BibitemShut
  {NoStop}%
\bibitem [{\citenamefont {Khamehchi}\ \emph {et~al.}(2014)\citenamefont
  {Khamehchi}, \citenamefont {Zhang}, \citenamefont {Hamner}, \citenamefont
  {Busch},\ and\ \citenamefont {Engels}}]{PhysRevA.90.063624}%
  \BibitemOpen
  \bibfield  {author} {\bibinfo {author} {\bibfnamefont {M.~A.}\ \bibnamefont
  {Khamehchi}}, \bibinfo {author} {\bibfnamefont {Y.}~\bibnamefont {Zhang}},
  \bibinfo {author} {\bibfnamefont {C.}~\bibnamefont {Hamner}}, \bibinfo
  {author} {\bibfnamefont {T.}~\bibnamefont {Busch}}, \ and\ \bibinfo {author}
  {\bibfnamefont {P.}~\bibnamefont {Engels}},\ }\href {\doibase
  10.1103/PhysRevA.90.063624} {\bibfield  {journal} {\bibinfo  {journal} {Phys.
  Rev. A}\ }\textbf {\bibinfo {volume} {90}},\ \bibinfo {pages} {063624}
  (\bibinfo {year} {2014})}\BibitemShut {NoStop}%
\bibitem [{\citenamefont {Ji}\ \emph {et~al.}(2015)\citenamefont {Ji},
  \citenamefont {Zhang}, \citenamefont {Xu}, \citenamefont {Wu}, \citenamefont
  {Deng}, \citenamefont {Chen},\ and\ \citenamefont
  {Pan}}]{PhysRevLett.114.105301}%
  \BibitemOpen
  \bibfield  {author} {\bibinfo {author} {\bibfnamefont {S.-C.}\ \bibnamefont
  {Ji}}, \bibinfo {author} {\bibfnamefont {L.}~\bibnamefont {Zhang}}, \bibinfo
  {author} {\bibfnamefont {X.-T.}\ \bibnamefont {Xu}}, \bibinfo {author}
  {\bibfnamefont {Z.}~\bibnamefont {Wu}}, \bibinfo {author} {\bibfnamefont
  {Y.}~\bibnamefont {Deng}}, \bibinfo {author} {\bibfnamefont {S.}~\bibnamefont
  {Chen}}, \ and\ \bibinfo {author} {\bibfnamefont {J.-W.}\ \bibnamefont
  {Pan}},\ }\href {\doibase 10.1103/PhysRevLett.114.105301} {\bibfield
  {journal} {\bibinfo  {journal} {Phys. Rev. Lett.}\ }\textbf {\bibinfo
  {volume} {114}},\ \bibinfo {pages} {105301} (\bibinfo {year}
  {2015})}\BibitemShut {NoStop}%
\bibitem [{\citenamefont {Tylutki}\ \emph {et~al.}(2020)\citenamefont
  {Tylutki}, \citenamefont {Astrakharchik}, \citenamefont {Malomed},\ and\
  \citenamefont {Petrov}}]{PhysRevA.101.051601}%
  \BibitemOpen
  \bibfield  {author} {\bibinfo {author} {\bibfnamefont {M.}~\bibnamefont
  {Tylutki}}, \bibinfo {author} {\bibfnamefont {G.~E.}\ \bibnamefont
  {Astrakharchik}}, \bibinfo {author} {\bibfnamefont {B.~A.}\ \bibnamefont
  {Malomed}}, \ and\ \bibinfo {author} {\bibfnamefont {D.~S.}\ \bibnamefont
  {Petrov}},\ }\href {\doibase 10.1103/PhysRevA.101.051601} {\bibfield
  {journal} {\bibinfo  {journal} {Phys. Rev. A}\ }\textbf {\bibinfo {volume}
  {101}},\ \bibinfo {pages} {051601} (\bibinfo {year} {2020})}\BibitemShut
  {NoStop}%
\bibitem [{\citenamefont {Isoshima}\ \emph {et~al.}(2000)\citenamefont
  {Isoshima}, \citenamefont {Ohmi},\ and\ \citenamefont
  {Machida}}]{JPSJ.69.3864}%
  \BibitemOpen
  \bibfield  {author} {\bibinfo {author} {\bibfnamefont {T.}~\bibnamefont
  {Isoshima}}, \bibinfo {author} {\bibfnamefont {T.}~\bibnamefont {Ohmi}}, \
  and\ \bibinfo {author} {\bibfnamefont {K.}~\bibnamefont {Machida}},\ }\href
  {\doibase 10.1143/JPSJ.69.3864} {\bibfield  {journal} {\bibinfo  {journal}
  {J. Phy. Soc. Jpn}\ }\textbf {\bibinfo {volume} {69}},\ \bibinfo {pages}
  {3864} (\bibinfo {year} {2000})}\BibitemShut {NoStop}%
\bibitem [{\citenamefont {Zhang}\ \emph {et~al.}(2004)\citenamefont {Zhang},
  \citenamefont {Yi},\ and\ \citenamefont {You}}]{PhysRevA.70.043611}%
  \BibitemOpen
  \bibfield  {author} {\bibinfo {author} {\bibfnamefont {W.}~\bibnamefont
  {Zhang}}, \bibinfo {author} {\bibfnamefont {S.}~\bibnamefont {Yi}}, \ and\
  \bibinfo {author} {\bibfnamefont {L.}~\bibnamefont {You}},\ }\href {\doibase
  10.1103/PhysRevA.70.043611} {\bibfield  {journal} {\bibinfo  {journal} {Phys.
  Rev. A}\ }\textbf {\bibinfo {volume} {70}},\ \bibinfo {pages} {043611}
  (\bibinfo {year} {2004})}\BibitemShut {NoStop}%
\bibitem [{\citenamefont {Phuc}\ \emph {et~al.}(2011)\citenamefont {Phuc},
  \citenamefont {Kawaguchi},\ and\ \citenamefont {Ueda}}]{PhysRevA.84.043645}%
  \BibitemOpen
  \bibfield  {author} {\bibinfo {author} {\bibfnamefont {N.~T.}\ \bibnamefont
  {Phuc}}, \bibinfo {author} {\bibfnamefont {Y.}~\bibnamefont {Kawaguchi}}, \
  and\ \bibinfo {author} {\bibfnamefont {M.}~\bibnamefont {Ueda}},\ }\href
  {\doibase 10.1103/PhysRevA.84.043645} {\bibfield  {journal} {\bibinfo
  {journal} {Phys. Rev. A}\ }\textbf {\bibinfo {volume} {84}},\ \bibinfo
  {pages} {043645} (\bibinfo {year} {2011})}\BibitemShut {NoStop}%
\bibitem [{\citenamefont {Kawaguchi}\ \emph {et~al.}(2012)\citenamefont
  {Kawaguchi}, \citenamefont {Phuc},\ and\ \citenamefont
  {Blakie}}]{PhysRevA.85.053611}%
  \BibitemOpen
  \bibfield  {author} {\bibinfo {author} {\bibfnamefont {Y.}~\bibnamefont
  {Kawaguchi}}, \bibinfo {author} {\bibfnamefont {N.~T.}\ \bibnamefont {Phuc}},
  \ and\ \bibinfo {author} {\bibfnamefont {P.~B.}\ \bibnamefont {Blakie}},\
  }\href {\doibase 10.1103/PhysRevA.85.053611} {\bibfield  {journal} {\bibinfo
  {journal} {Phys. Rev. A}\ }\textbf {\bibinfo {volume} {85}},\ \bibinfo
  {pages} {053611} (\bibinfo {year} {2012})}\BibitemShut {NoStop}%
\bibitem [{\citenamefont {Jacob}\ \emph {et~al.}(2012)\citenamefont {Jacob},
  \citenamefont {Shao}, \citenamefont {Corre}, \citenamefont {Zibold},
  \citenamefont {De~Sarlo}, \citenamefont {Mimoun}, \citenamefont {Dalibard},\
  and\ \citenamefont {Gerbier}}]{PhysRevA.86.061601}%
  \BibitemOpen
  \bibfield  {author} {\bibinfo {author} {\bibfnamefont {D.}~\bibnamefont
  {Jacob}}, \bibinfo {author} {\bibfnamefont {L.}~\bibnamefont {Shao}},
  \bibinfo {author} {\bibfnamefont {V.}~\bibnamefont {Corre}}, \bibinfo
  {author} {\bibfnamefont {T.}~\bibnamefont {Zibold}}, \bibinfo {author}
  {\bibfnamefont {L.}~\bibnamefont {De~Sarlo}}, \bibinfo {author}
  {\bibfnamefont {E.}~\bibnamefont {Mimoun}}, \bibinfo {author} {\bibfnamefont
  {J.}~\bibnamefont {Dalibard}}, \ and\ \bibinfo {author} {\bibfnamefont
  {F.}~\bibnamefont {Gerbier}},\ }\href {\doibase 10.1103/PhysRevA.86.061601}
  {\bibfield  {journal} {\bibinfo  {journal} {Phys. Rev. A}\ }\textbf {\bibinfo
  {volume} {86}},\ \bibinfo {pages} {061601} (\bibinfo {year}
  {2012})}\BibitemShut {NoStop}%
\bibitem [{\citenamefont {Mur-Petit}\ \emph {et~al.}(2006)\citenamefont
  {Mur-Petit}, \citenamefont {Guilleumas}, \citenamefont {Polls}, \citenamefont
  {Sanpera}, \citenamefont {Lewenstein}, \citenamefont {Bongs},\ and\
  \citenamefont {Sengstock}}]{PhysRevA.73.013629}%
  \BibitemOpen
  \bibfield  {author} {\bibinfo {author} {\bibfnamefont {J.}~\bibnamefont
  {Mur-Petit}}, \bibinfo {author} {\bibfnamefont {M.}~\bibnamefont
  {Guilleumas}}, \bibinfo {author} {\bibfnamefont {A.}~\bibnamefont {Polls}},
  \bibinfo {author} {\bibfnamefont {A.}~\bibnamefont {Sanpera}}, \bibinfo
  {author} {\bibfnamefont {M.}~\bibnamefont {Lewenstein}}, \bibinfo {author}
  {\bibfnamefont {K.}~\bibnamefont {Bongs}}, \ and\ \bibinfo {author}
  {\bibfnamefont {K.}~\bibnamefont {Sengstock}},\ }\href {\doibase
  10.1103/PhysRevA.73.013629} {\bibfield  {journal} {\bibinfo  {journal} {Phys.
  Rev. A}\ }\textbf {\bibinfo {volume} {73}},\ \bibinfo {pages} {013629}
  (\bibinfo {year} {2006})}\BibitemShut {NoStop}%
\bibitem [{\citenamefont {Moreno-Cardoner}\ \emph {et~al.}(2007)\citenamefont
  {Moreno-Cardoner}, \citenamefont {Mur-Petit}, \citenamefont {Guilleumas},
  \citenamefont {Polls}, \citenamefont {Sanpera},\ and\ \citenamefont
  {Lewenstein}}]{PhysRevLett.99.020404}%
  \BibitemOpen
  \bibfield  {author} {\bibinfo {author} {\bibfnamefont {M.}~\bibnamefont
  {Moreno-Cardoner}}, \bibinfo {author} {\bibfnamefont {J.}~\bibnamefont
  {Mur-Petit}}, \bibinfo {author} {\bibfnamefont {M.}~\bibnamefont
  {Guilleumas}}, \bibinfo {author} {\bibfnamefont {A.}~\bibnamefont {Polls}},
  \bibinfo {author} {\bibfnamefont {A.}~\bibnamefont {Sanpera}}, \ and\
  \bibinfo {author} {\bibfnamefont {M.}~\bibnamefont {Lewenstein}},\ }\href
  {\doibase 10.1103/PhysRevLett.99.020404} {\bibfield  {journal} {\bibinfo
  {journal} {Phys. Rev. Lett.}\ }\textbf {\bibinfo {volume} {99}},\ \bibinfo
  {pages} {020404} (\bibinfo {year} {2007})}\BibitemShut {NoStop}%
\bibitem [{\citenamefont {Witkowska}\ \emph {et~al.}(2014)\citenamefont
  {Witkowska}, \citenamefont {\ifmmode~\acute{S}\else \'{S}\fi{}wis\l{}ocki},\
  and\ \citenamefont {Matuszewski}}]{PhysRevA.90.033604}%
  \BibitemOpen
  \bibfield  {author} {\bibinfo {author} {\bibfnamefont {E.}~\bibnamefont
  {Witkowska}}, \bibinfo {author} {\bibfnamefont {T.}~\bibnamefont
  {\ifmmode~\acute{S}\else \'{S}\fi{}wis\l{}ocki}}, \ and\ \bibinfo {author}
  {\bibfnamefont {M.}~\bibnamefont {Matuszewski}},\ }\href {\doibase
  10.1103/PhysRevA.90.033604} {\bibfield  {journal} {\bibinfo  {journal} {Phys.
  Rev. A}\ }\textbf {\bibinfo {volume} {90}},\ \bibinfo {pages} {033604}
  (\bibinfo {year} {2014})}\BibitemShut {NoStop}%
\bibitem [{\citenamefont {Chen}\ \emph
  {et~al.}(2017{\natexlab{b}})\citenamefont {Chen}, \citenamefont {Liu},\ and\
  \citenamefont {Hu}}]{PhysRevA.96.013625}%
  \BibitemOpen
  \bibfield  {author} {\bibinfo {author} {\bibfnamefont {X.-L.}\ \bibnamefont
  {Chen}}, \bibinfo {author} {\bibfnamefont {X.-J.}\ \bibnamefont {Liu}}, \
  and\ \bibinfo {author} {\bibfnamefont {H.}~\bibnamefont {Hu}},\ }\href
  {\doibase 10.1103/PhysRevA.96.013625} {\bibfield  {journal} {\bibinfo
  {journal} {Phys. Rev. A}\ }\textbf {\bibinfo {volume} {96}},\ \bibinfo
  {pages} {013625} (\bibinfo {year} {2017}{\natexlab{b}})}\BibitemShut
  {NoStop}%
\bibitem [{\citenamefont {Ji}\ \emph {et~al.}(2014)\citenamefont {Ji},
  \citenamefont {Zhang}, \citenamefont {Zhang}, \citenamefont {Du},
  \citenamefont {Zheng}, \citenamefont {Deng}, \citenamefont {Zhai},
  \citenamefont {Chen},\ and\ \citenamefont {Pan}}]{ji2014experimental}%
  \BibitemOpen
  \bibfield  {author} {\bibinfo {author} {\bibfnamefont {S.-C.}\ \bibnamefont
  {Ji}}, \bibinfo {author} {\bibfnamefont {J.-Y.}\ \bibnamefont {Zhang}},
  \bibinfo {author} {\bibfnamefont {L.}~\bibnamefont {Zhang}}, \bibinfo
  {author} {\bibfnamefont {Z.-D.}\ \bibnamefont {Du}}, \bibinfo {author}
  {\bibfnamefont {W.}~\bibnamefont {Zheng}}, \bibinfo {author} {\bibfnamefont
  {Y.-J.}\ \bibnamefont {Deng}}, \bibinfo {author} {\bibfnamefont
  {H.}~\bibnamefont {Zhai}}, \bibinfo {author} {\bibfnamefont {S.}~\bibnamefont
  {Chen}}, \ and\ \bibinfo {author} {\bibfnamefont {J.-W.}\ \bibnamefont
  {Pan}},\ }\href {\doibase 10.1038/nphys2905} {\bibfield  {journal} {\bibinfo
  {journal} {Nat. Phys.}\ }\textbf {\bibinfo {volume} {10}},\ \bibinfo {pages}
  {314} (\bibinfo {year} {2014})}\BibitemShut {NoStop}%
\bibitem [{\citenamefont {Yu}(2014)}]{PhysRevA.90.053608}%
  \BibitemOpen
  \bibfield  {author} {\bibinfo {author} {\bibfnamefont {Z.-Q.}\ \bibnamefont
  {Yu}},\ }\href {\doibase 10.1103/PhysRevA.90.053608} {\bibfield  {journal}
  {\bibinfo  {journal} {Phys. Rev. A}\ }\textbf {\bibinfo {volume} {90}},\
  \bibinfo {pages} {053608} (\bibinfo {year} {2014})}\BibitemShut {NoStop}%
\bibitem [{\citenamefont {Ozawa}\ and\ \citenamefont
  {Baym}(2012)}]{PhysRevLett.109.025301}%
  \BibitemOpen
  \bibfield  {author} {\bibinfo {author} {\bibfnamefont {T.}~\bibnamefont
  {Ozawa}}\ and\ \bibinfo {author} {\bibfnamefont {G.}~\bibnamefont {Baym}},\
  }\href {\doibase 10.1103/PhysRevLett.109.025301} {\bibfield  {journal}
  {\bibinfo  {journal} {Phys. Rev. Lett.}\ }\textbf {\bibinfo {volume} {109}},\
  \bibinfo {pages} {025301} (\bibinfo {year} {2012})}\BibitemShut {NoStop}%
\bibitem [{\citenamefont {Kawasaki}\ and\ \citenamefont
  {Holzmann}(2017)}]{PhysRevA.95.051601}%
  \BibitemOpen
  \bibfield  {author} {\bibinfo {author} {\bibfnamefont {E.}~\bibnamefont
  {Kawasaki}}\ and\ \bibinfo {author} {\bibfnamefont {M.}~\bibnamefont
  {Holzmann}},\ }\href {\doibase 10.1103/PhysRevA.95.051601} {\bibfield
  {journal} {\bibinfo  {journal} {Phys. Rev. A}\ }\textbf {\bibinfo {volume}
  {95}},\ \bibinfo {pages} {051601} (\bibinfo {year} {2017})}\BibitemShut
  {NoStop}%
\bibitem [{\citenamefont {Su}\ \emph {et~al.}(2017)\citenamefont {Su},
  \citenamefont {Liu}, \citenamefont {Gou}, \citenamefont {Liao}, \citenamefont
  {Fialko},\ and\ \citenamefont {Brand}}]{PhysRevA.95.053629}%
  \BibitemOpen
  \bibfield  {author} {\bibinfo {author} {\bibfnamefont {S.-W.}\ \bibnamefont
  {Su}}, \bibinfo {author} {\bibfnamefont {I.-K.}\ \bibnamefont {Liu}},
  \bibinfo {author} {\bibfnamefont {S.-C.}\ \bibnamefont {Gou}}, \bibinfo
  {author} {\bibfnamefont {R.}~\bibnamefont {Liao}}, \bibinfo {author}
  {\bibfnamefont {O.}~\bibnamefont {Fialko}}, \ and\ \bibinfo {author}
  {\bibfnamefont {J.}~\bibnamefont {Brand}},\ }\href {\doibase
  10.1103/PhysRevA.95.053629} {\bibfield  {journal} {\bibinfo  {journal} {Phys.
  Rev. A}\ }\textbf {\bibinfo {volume} {95}},\ \bibinfo {pages} {053629}
  (\bibinfo {year} {2017})}\BibitemShut {NoStop}%
\bibitem [{\citenamefont {Wilson}\ \emph {et~al.}(2010)\citenamefont {Wilson},
  \citenamefont {Ronen},\ and\ \citenamefont {Bohn}}]{PhysRevLett.104.094501}%
  \BibitemOpen
  \bibfield  {author} {\bibinfo {author} {\bibfnamefont {R.~M.}\ \bibnamefont
  {Wilson}}, \bibinfo {author} {\bibfnamefont {S.}~\bibnamefont {Ronen}}, \
  and\ \bibinfo {author} {\bibfnamefont {J.~L.}\ \bibnamefont {Bohn}},\ }\href
  {\doibase 10.1103/PhysRevLett.104.094501} {\bibfield  {journal} {\bibinfo
  {journal} {Phys. Rev. Lett.}\ }\textbf {\bibinfo {volume} {104}},\ \bibinfo
  {pages} {094501} (\bibinfo {year} {2010})}\BibitemShut {NoStop}%
\bibitem [{\citenamefont {Ho}(1998)}]{PhysRevLett.81.742}%
  \BibitemOpen
  \bibfield  {author} {\bibinfo {author} {\bibfnamefont {T.-L.}\ \bibnamefont
  {Ho}},\ }\href {\doibase 10.1103/PhysRevLett.81.742} {\bibfield  {journal}
  {\bibinfo  {journal} {Phys. Rev. Lett.}\ }\textbf {\bibinfo {volume} {81}},\
  \bibinfo {pages} {742} (\bibinfo {year} {1998})}\BibitemShut {NoStop}%
\bibitem [{\citenamefont {Ohmi}\ and\ \citenamefont
  {Machida}(1998)}]{JPSJ.67.1822}%
  \BibitemOpen
  \bibfield  {author} {\bibinfo {author} {\bibfnamefont {T.}~\bibnamefont
  {Ohmi}}\ and\ \bibinfo {author} {\bibfnamefont {K.}~\bibnamefont {Machida}},\
  }\href {\doibase 10.1143/JPSJ.67.1822} {\bibfield  {journal} {\bibinfo
  {journal} {J. Phy. Soc. Jpn}\ }\textbf {\bibinfo {volume} {67}},\ \bibinfo
  {pages} {1822} (\bibinfo {year} {1998})}\BibitemShut {NoStop}%
\bibitem [{\citenamefont {Griffin}(1996)}]{PhysRevB.53.9341}%
  \BibitemOpen
  \bibfield  {author} {\bibinfo {author} {\bibfnamefont {A.}~\bibnamefont
  {Griffin}},\ }\href {\doibase 10.1103/PhysRevB.53.9341} {\bibfield  {journal}
  {\bibinfo  {journal} {Phys. Rev. B}\ }\textbf {\bibinfo {volume} {53}},\
  \bibinfo {pages} {9341} (\bibinfo {year} {1996})}\BibitemShut {NoStop}%
\bibitem [{\citenamefont {Roy}\ \emph {et~al.}(2014)\citenamefont {Roy},
  \citenamefont {Gautam},\ and\ \citenamefont {Angom}}]{PhysRevA.89.013617}%
  \BibitemOpen
  \bibfield  {author} {\bibinfo {author} {\bibfnamefont {A.}~\bibnamefont
  {Roy}}, \bibinfo {author} {\bibfnamefont {S.}~\bibnamefont {Gautam}}, \ and\
  \bibinfo {author} {\bibfnamefont {D.}~\bibnamefont {Angom}},\ }\href
  {\doibase 10.1103/PhysRevA.89.013617} {\bibfield  {journal} {\bibinfo
  {journal} {Phys. Rev. A}\ }\textbf {\bibinfo {volume} {89}},\ \bibinfo
  {pages} {013617} (\bibinfo {year} {2014})}\BibitemShut {NoStop}%
\bibitem [{\citenamefont {Roy}\ and\ \citenamefont
  {Angom}(2014)}]{PhysRevA.90.023612}%
  \BibitemOpen
  \bibfield  {author} {\bibinfo {author} {\bibfnamefont {A.}~\bibnamefont
  {Roy}}\ and\ \bibinfo {author} {\bibfnamefont {D.}~\bibnamefont {Angom}},\
  }\href {\doibase 10.1103/PhysRevA.90.023612} {\bibfield  {journal} {\bibinfo
  {journal} {Phys. Rev. A}\ }\textbf {\bibinfo {volume} {90}},\ \bibinfo
  {pages} {023612} (\bibinfo {year} {2014})}\BibitemShut {NoStop}%
\bibitem [{\citenamefont {Hugenholtz}\ and\ \citenamefont
  {Pines}(1959)}]{PhysRev.116.489}%
  \BibitemOpen
  \bibfield  {author} {\bibinfo {author} {\bibfnamefont {N.~M.}\ \bibnamefont
  {Hugenholtz}}\ and\ \bibinfo {author} {\bibfnamefont {D.}~\bibnamefont
  {Pines}},\ }\href {\doibase 10.1103/PhysRev.116.489} {\bibfield  {journal}
  {\bibinfo  {journal} {Phys. Rev.}\ }\textbf {\bibinfo {volume} {116}},\
  \bibinfo {pages} {489} (\bibinfo {year} {1959})}\BibitemShut {NoStop}%
\bibitem [{\citenamefont {Kaur}\ \emph {et~al.}(2021)\citenamefont {Kaur},
  \citenamefont {Roy},\ and\ \citenamefont {Gautam}}]{KAUR2021107671}%
  \BibitemOpen
  \bibfield  {author} {\bibinfo {author} {\bibfnamefont {P.}~\bibnamefont
  {Kaur}}, \bibinfo {author} {\bibfnamefont {A.}~\bibnamefont {Roy}}, \ and\
  \bibinfo {author} {\bibfnamefont {S.}~\bibnamefont {Gautam}},\ }\href
  {\doibase https://doi.org/10.1016/j.cpc.2020.107671} {\bibfield  {journal}
  {\bibinfo  {journal} {Comput. Phys. Commun.}\ }\textbf {\bibinfo {volume}
  {259}},\ \bibinfo {pages} {107671} (\bibinfo {year} {2021})}\BibitemShut
  {NoStop}%
\bibitem [{\citenamefont {Banger}\ \emph {et~al.}(2021)\citenamefont {Banger},
  \citenamefont {Kaur},\ and\ \citenamefont {Gautam}}]{banger2021semi}%
  \BibitemOpen
  \bibfield  {author} {\bibinfo {author} {\bibfnamefont {P.}~\bibnamefont
  {Banger}}, \bibinfo {author} {\bibfnamefont {P.}~\bibnamefont {Kaur}}, \ and\
  \bibinfo {author} {\bibfnamefont {S.}~\bibnamefont {Gautam}},\ }\href
  {\doibase 10.1142/S0129183122500462} {\bibfield  {journal} {\bibinfo
  {journal} {Int. J. Mod. Phys. C}\ ,\ \bibinfo {pages} {2250046}} (\bibinfo
  {year} {2021})}\BibitemShut {NoStop}%
\bibitem [{\citenamefont {Gao}\ and\ \citenamefont
  {Cai}(2020)}]{GAO2020109058}%
  \BibitemOpen
  \bibfield  {author} {\bibinfo {author} {\bibfnamefont {Y.}~\bibnamefont
  {Gao}}\ and\ \bibinfo {author} {\bibfnamefont {Y.}~\bibnamefont {Cai}},\
  }\href {\doibase https://doi.org/10.1016/j.jcp.2019.109058} {\bibfield
  {journal} {\bibinfo  {journal} {J. Comput. Phys.}\ }\textbf {\bibinfo
  {volume} {403}},\ \bibinfo {pages} {109058} (\bibinfo {year}
  {2020})}\BibitemShut {NoStop}%
\bibitem [{\citenamefont {Serrano-Ens\'astiga}\ and\ \citenamefont
  {Mireles}(2021)}]{PhysRevA.104.063308}%
  \BibitemOpen
  \bibfield  {author} {\bibinfo {author} {\bibfnamefont {E.}~\bibnamefont
  {Serrano-Ens\'astiga}}\ and\ \bibinfo {author} {\bibfnamefont
  {F.}~\bibnamefont {Mireles}},\ }\href {\doibase 10.1103/PhysRevA.104.063308}
  {\bibfield  {journal} {\bibinfo  {journal} {Phys. Rev. A}\ }\textbf {\bibinfo
  {volume} {104}},\ \bibinfo {pages} {063308} (\bibinfo {year}
  {2021})}\BibitemShut {NoStop}%
\bibitem [{\citenamefont {Ticknor}(2014)}]{PhysRevA.89.053601}%
  \BibitemOpen
  \bibfield  {author} {\bibinfo {author} {\bibfnamefont {C.}~\bibnamefont
  {Ticknor}},\ }\href {\doibase 10.1103/PhysRevA.89.053601} {\bibfield
  {journal} {\bibinfo  {journal} {Phys. Rev. A}\ }\textbf {\bibinfo {volume}
  {89}},\ \bibinfo {pages} {053601} (\bibinfo {year} {2014})}\BibitemShut
  {NoStop}%
\bibitem [{\citenamefont {Pal}\ \emph {et~al.}(2018)\citenamefont {Pal},
  \citenamefont {Roy},\ and\ \citenamefont {Angom}}]{Pal_2018}%
  \BibitemOpen
  \bibfield  {author} {\bibinfo {author} {\bibfnamefont {S.}~\bibnamefont
  {Pal}}, \bibinfo {author} {\bibfnamefont {A.}~\bibnamefont {Roy}}, \ and\
  \bibinfo {author} {\bibfnamefont {D.}~\bibnamefont {Angom}},\ }\href
  {\doibase 10.1088/1361-6455/aab541} {\bibfield  {journal} {\bibinfo
  {journal} {J. Phys. B: At. Mol. Opt. Phys.}\ }\textbf {\bibinfo {volume}
  {51}},\ \bibinfo {pages} {085302} (\bibinfo {year} {2018})}\BibitemShut
  {NoStop}%
\bibitem [{\citenamefont {Knoop}\ \emph {et~al.}(2011)\citenamefont {Knoop},
  \citenamefont {Schuster}, \citenamefont {Scelle}, \citenamefont {Trautmann},
  \citenamefont {Appmeier}, \citenamefont {Oberthaler}, \citenamefont
  {Tiesinga},\ and\ \citenamefont {Tiemann}}]{PhysRevA.83.042704}%
  \BibitemOpen
  \bibfield  {author} {\bibinfo {author} {\bibfnamefont {S.}~\bibnamefont
  {Knoop}}, \bibinfo {author} {\bibfnamefont {T.}~\bibnamefont {Schuster}},
  \bibinfo {author} {\bibfnamefont {R.}~\bibnamefont {Scelle}}, \bibinfo
  {author} {\bibfnamefont {A.}~\bibnamefont {Trautmann}}, \bibinfo {author}
  {\bibfnamefont {J.}~\bibnamefont {Appmeier}}, \bibinfo {author}
  {\bibfnamefont {M.~K.}\ \bibnamefont {Oberthaler}}, \bibinfo {author}
  {\bibfnamefont {E.}~\bibnamefont {Tiesinga}}, \ and\ \bibinfo {author}
  {\bibfnamefont {E.}~\bibnamefont {Tiemann}},\ }\href {\doibase
  10.1103/PhysRevA.83.042704} {\bibfield  {journal} {\bibinfo  {journal} {Phys.
  Rev. A}\ }\textbf {\bibinfo {volume} {83}},\ \bibinfo {pages} {042704}
  (\bibinfo {year} {2011})}\BibitemShut {NoStop}%
\bibitem [{\citenamefont {Watanabe}\ and\ \citenamefont
  {Murayama}(2012)}]{PhysRevLett.108.251602}%
  \BibitemOpen
  \bibfield  {author} {\bibinfo {author} {\bibfnamefont {H.}~\bibnamefont
  {Watanabe}}\ and\ \bibinfo {author} {\bibfnamefont {H.}~\bibnamefont
  {Murayama}},\ }\href {\doibase 10.1103/PhysRevLett.108.251602} {\bibfield
  {journal} {\bibinfo  {journal} {Phys. Rev. Lett.}\ }\textbf {\bibinfo
  {volume} {108}},\ \bibinfo {pages} {251602} (\bibinfo {year}
  {2012})}\BibitemShut {NoStop}%
\bibitem [{\citenamefont {Dobson}(1994)}]{PhysRevLett.73.2244}%
  \BibitemOpen
  \bibfield  {author} {\bibinfo {author} {\bibfnamefont {J.~F.}\ \bibnamefont
  {Dobson}},\ }\href {\doibase 10.1103/PhysRevLett.73.2244} {\bibfield
  {journal} {\bibinfo  {journal} {Phys. Rev. Lett.}\ }\textbf {\bibinfo
  {volume} {73}},\ \bibinfo {pages} {2244} (\bibinfo {year}
  {1994})}\BibitemShut {NoStop}%
\bibitem [{\citenamefont {Bienaim\'e}\ \emph {et~al.}(2016)\citenamefont
  {Bienaim\'e}, \citenamefont {Fava}, \citenamefont {Colzi}, \citenamefont
  {Mordini}, \citenamefont {Serafini}, \citenamefont {Qu}, \citenamefont
  {Stringari}, \citenamefont {Lamporesi},\ and\ \citenamefont
  {Ferrari}}]{PhysRevA.94.063652}%
  \BibitemOpen
  \bibfield  {author} {\bibinfo {author} {\bibfnamefont {T.}~\bibnamefont
  {Bienaim\'e}}, \bibinfo {author} {\bibfnamefont {E.}~\bibnamefont {Fava}},
  \bibinfo {author} {\bibfnamefont {G.}~\bibnamefont {Colzi}}, \bibinfo
  {author} {\bibfnamefont {C.}~\bibnamefont {Mordini}}, \bibinfo {author}
  {\bibfnamefont {S.}~\bibnamefont {Serafini}}, \bibinfo {author}
  {\bibfnamefont {C.}~\bibnamefont {Qu}}, \bibinfo {author} {\bibfnamefont
  {S.}~\bibnamefont {Stringari}}, \bibinfo {author} {\bibfnamefont
  {G.}~\bibnamefont {Lamporesi}}, \ and\ \bibinfo {author} {\bibfnamefont
  {G.}~\bibnamefont {Ferrari}},\ }\href {\doibase 10.1103/PhysRevA.94.063652}
  {\bibfield  {journal} {\bibinfo  {journal} {Phys. Rev. A}\ }\textbf {\bibinfo
  {volume} {94}},\ \bibinfo {pages} {063652} (\bibinfo {year}
  {2016})}\BibitemShut {NoStop}%
\bibitem [{\citenamefont {P\'erez-Garc\'{\i}a}\ \emph
  {et~al.}(1996)\citenamefont {P\'erez-Garc\'{\i}a}, \citenamefont {Michinel},
  \citenamefont {Cirac}, \citenamefont {Lewenstein},\ and\ \citenamefont
  {Zoller}}]{PhysRevLett.77.5320}%
  \BibitemOpen
  \bibfield  {author} {\bibinfo {author} {\bibfnamefont {V.~M.}\ \bibnamefont
  {P\'erez-Garc\'{\i}a}}, \bibinfo {author} {\bibfnamefont {H.}~\bibnamefont
  {Michinel}}, \bibinfo {author} {\bibfnamefont {J.~I.}\ \bibnamefont {Cirac}},
  \bibinfo {author} {\bibfnamefont {M.}~\bibnamefont {Lewenstein}}, \ and\
  \bibinfo {author} {\bibfnamefont {P.}~\bibnamefont {Zoller}},\ }\href
  {\doibase 10.1103/PhysRevLett.77.5320} {\bibfield  {journal} {\bibinfo
  {journal} {Phys. Rev. Lett.}\ }\textbf {\bibinfo {volume} {77}},\ \bibinfo
  {pages} {5320} (\bibinfo {year} {1996})}\BibitemShut {NoStop}%
\bibitem [{\citenamefont {P\'erez-Garc\'{\i}a}\ \emph
  {et~al.}(1997)\citenamefont {P\'erez-Garc\'{\i}a}, \citenamefont {Michinel},
  \citenamefont {Cirac}, \citenamefont {Lewenstein},\ and\ \citenamefont
  {Zoller}}]{PhysRevA.56.1424}%
  \BibitemOpen
  \bibfield  {author} {\bibinfo {author} {\bibfnamefont {V.~M.}\ \bibnamefont
  {P\'erez-Garc\'{\i}a}}, \bibinfo {author} {\bibfnamefont {H.}~\bibnamefont
  {Michinel}}, \bibinfo {author} {\bibfnamefont {J.~I.}\ \bibnamefont {Cirac}},
  \bibinfo {author} {\bibfnamefont {M.}~\bibnamefont {Lewenstein}}, \ and\
  \bibinfo {author} {\bibfnamefont {P.}~\bibnamefont {Zoller}},\ }\href
  {\doibase 10.1103/PhysRevA.56.1424} {\bibfield  {journal} {\bibinfo
  {journal} {Phys. Rev. A}\ }\textbf {\bibinfo {volume} {56}},\ \bibinfo
  {pages} {1424} (\bibinfo {year} {1997})}\BibitemShut {NoStop}%
\bibitem [{\citenamefont {Ketterle}\ and\ \citenamefont {van
  Druten}(1996)}]{PhysRevA.54.656}%
  \BibitemOpen
  \bibfield  {author} {\bibinfo {author} {\bibfnamefont {W.}~\bibnamefont
  {Ketterle}}\ and\ \bibinfo {author} {\bibfnamefont {N.~J.}\ \bibnamefont {van
  Druten}},\ }\href {\doibase 10.1103/PhysRevA.54.656} {\bibfield  {journal}
  {\bibinfo  {journal} {Phys. Rev. A}\ }\textbf {\bibinfo {volume} {54}},\
  \bibinfo {pages} {656} (\bibinfo {year} {1996})}\BibitemShut {NoStop}%
\bibitem [{\citenamefont {{Kao, Y.-M.}}\ and\ \citenamefont {{Jiang, T.
  F.}}(2006)}]{refId0}%
  \BibitemOpen
  \bibfield  {author} {\bibinfo {author} {\bibnamefont {{Kao, Y.-M.}}}\ and\
  \bibinfo {author} {\bibnamefont {{Jiang, T. F.}}},\ }\href {\doibase
  10.1140/epjd/e2006-00157-4} {\bibfield  {journal} {\bibinfo  {journal} {Eur.
  Phys. J. D}\ }\textbf {\bibinfo {volume} {40}},\ \bibinfo {pages} {263}
  (\bibinfo {year} {2006})}\BibitemShut {NoStop}%
\bibitem [{\citenamefont {Huang}\ \emph {et~al.}(2002)\citenamefont {Huang},
  \citenamefont {Gou},\ and\ \citenamefont {Tsai}}]{PhysRevA.65.063610}%
  \BibitemOpen
  \bibfield  {author} {\bibinfo {author} {\bibfnamefont {W.-J.}\ \bibnamefont
  {Huang}}, \bibinfo {author} {\bibfnamefont {S.-C.}\ \bibnamefont {Gou}}, \
  and\ \bibinfo {author} {\bibfnamefont {Y.-C.}\ \bibnamefont {Tsai}},\ }\href
  {\doibase 10.1103/PhysRevA.65.063610} {\bibfield  {journal} {\bibinfo
  {journal} {Phys. Rev. A}\ }\textbf {\bibinfo {volume} {65}},\ \bibinfo
  {pages} {063610} (\bibinfo {year} {2002})}\BibitemShut {NoStop}%
\bibitem [{\citenamefont {Gies}\ \emph {et~al.}(2004)\citenamefont {Gies},
  \citenamefont {van Zyl}, \citenamefont {Morgan},\ and\ \citenamefont
  {Hutchinson}}]{PhysRevA.69.023616}%
  \BibitemOpen
  \bibfield  {author} {\bibinfo {author} {\bibfnamefont {C.}~\bibnamefont
  {Gies}}, \bibinfo {author} {\bibfnamefont {B.~P.}\ \bibnamefont {van Zyl}},
  \bibinfo {author} {\bibfnamefont {S.~A.}\ \bibnamefont {Morgan}}, \ and\
  \bibinfo {author} {\bibfnamefont {D.~A.~W.}\ \bibnamefont {Hutchinson}},\
  }\href {\doibase 10.1103/PhysRevA.69.023616} {\bibfield  {journal} {\bibinfo
  {journal} {Phys. Rev. A}\ }\textbf {\bibinfo {volume} {69}},\ \bibinfo
  {pages} {023616} (\bibinfo {year} {2004})}\BibitemShut {NoStop}%
\bibitem [{\citenamefont {Dodd}\ \emph {et~al.}(1998)\citenamefont {Dodd},
  \citenamefont {Edwards}, \citenamefont {Clark},\ and\ \citenamefont
  {Burnett}}]{PhysRevA.57.R32}%
  \BibitemOpen
  \bibfield  {author} {\bibinfo {author} {\bibfnamefont {R.~J.}\ \bibnamefont
  {Dodd}}, \bibinfo {author} {\bibfnamefont {M.}~\bibnamefont {Edwards}},
  \bibinfo {author} {\bibfnamefont {C.~W.}\ \bibnamefont {Clark}}, \ and\
  \bibinfo {author} {\bibfnamefont {K.}~\bibnamefont {Burnett}},\ }\href
  {\doibase 10.1103/PhysRevA.57.R32} {\bibfield  {journal} {\bibinfo  {journal}
  {Phys. Rev. A}\ }\textbf {\bibinfo {volume} {57}},\ \bibinfo {pages} {R32}
  (\bibinfo {year} {1998})}\BibitemShut {NoStop}%
\bibitem [{\citenamefont {Shi}\ and\ \citenamefont
  {Zheng}(1999)}]{PhysRevA.59.1562}%
  \BibitemOpen
  \bibfield  {author} {\bibinfo {author} {\bibfnamefont {H.}~\bibnamefont
  {Shi}}\ and\ \bibinfo {author} {\bibfnamefont {W.-M.}\ \bibnamefont
  {Zheng}},\ }\href {\doibase 10.1103/PhysRevA.59.1562} {\bibfield  {journal}
  {\bibinfo  {journal} {Phys. Rev. A}\ }\textbf {\bibinfo {volume} {59}},\
  \bibinfo {pages} {1562} (\bibinfo {year} {1999})}\BibitemShut {NoStop}%
\end{thebibliography}%
\bibliographystyle{apsrev4-1}
\end{document}